\documentclass[12pt]{article} %***

\usepackage[sectionbib]{natbib}
\usepackage{array,epsfig,fancyhdr,rotating}
\usepackage{titlesec}
\usepackage{arydshln} 
\usepackage[]{hyperref}  %<----modified by Ivan
\usepackage{sectsty, secdot}
\sectionfont{\fontsize{12}{14pt plus.8pt minus .6pt}\selectfont}
\renewcommand{\theequation}{\thesection\arabic{equation}}
\subsectionfont{\fontsize{12}{14pt plus.8pt minus .6pt}\selectfont}
\usepackage{bm} % Bold font maths
\usepackage{makecell}
\RequirePackage{booktabs}
\usepackage{color} % Colour support
\usepackage{wrapfig} % Text flowing around figures
\usepackage{lipsum} % Generates meaningless text
\usepackage{enumitem}

\usepackage{multicol} % Multi Column Environment
\usepackage{multirow}
\usepackage{float} % Lets you add images to multi-column environment
\usepackage{diagbox}
\usepackage{amsmath}
\usepackage{amssymb}
\usepackage{amsfonts}
\usepackage{multirow}
\usepackage{amsthm}

\newtheorem{theorem}{Theorem}
\newtheorem{lemma}{Lemma}

\theoremstyle{definition}

\newcommand{\cD}{\mathcal{D}}
\newcommand{\cR}{\mathcal{R}}
\newcommand{\bbeta}{\bm \beta}
\newcommand{\bC}{\bm C}

\newcommand{\bM}{\bm M}
\newcommand{\bX}{\bm X}
\newcommand{\bY}{\bm Y}
\newcommand{\bbV}{\text{Var}}
\newcommand{\bbcov}{\text{cov}}
\begin{document}

%%%%%%%%%%%%%%%%%%%%%%%%%%%%%%%%%%%%%%%%%%%%%%%%%%%%%%%%%%%%%%%%%%%%%%%%%%%%%%%%%%%%%%%%%%%%%%%%%%%%%%%%%%%%%%%%%%%%%%%%%%%%
%%%%%%%%%%%%%%%%%%%%%%%%%%%%%%%%%%%%%%%%%%%%%%%%%%%%%%%%%%%%%%%%%%%%%%%%%%%%%%%%%%%%%%%%%%%%%%%%%%%%%%%%%%%%%%%%%%%%%%%%%%%%
% 设置行间距
\renewcommand{\baselinestretch}{1.5}
\renewcommand{\thefootnote}{}
$\ $\par

%%%%%%%%%%%%%%%%%%%%%%%%%%%%%%%%%%%%%%%%%%%%%%%%%%%%%%%%%%%%%%%%%%%%%%%%%%%%%%%%%%%%%%%%%%%%%%%%%%%%%%%%%%%%%%%%%%%%%%%%%%%%

\fontsize{12}{14pt plus.8pt minus .6pt}\selectfont \vspace{0.8pc}
\centerline{\large\bf TOTAL-EFFECT TEST MAY ERRONEOUSLY REJECT}
\vspace{2pt} 
\centerline{\large\bf SO-CALLED ``FULL" OR ``COMPLETE" MEDIATION}
% \centerline{\large\bf PROOFS AND EXAMPLES}
% \centerline{\large\bf }
\vspace{.4cm} 

\centerline{Tingxuan Han$^1$, Luxi Zhang$^2$, Xinshu Zhao$^2$ and Ke Deng$^1$} 
\vspace{.4cm} 
\centerline{\it $^1$Tsinghua University, $^2$University of Macau}
 \vspace{.55cm} \fontsize{9}{11.5pt plus.8pt minus.6pt} \selectfont

% 交叉索引
\let\nofiles\relax
%%%%%%%%%%%%%%%%%%%%%%%%%%%%%%%%%%%%%%%%%%%%%%%%%%%%%%%%%%%%%%%%%%%%%%%%%%%%%%%%%%%%%%%%%%%%%%%%%%%%%%%%%%%%%%%%%%%%%%%%%%%%

\begin{quotation}
\noindent {\it Abstract}.  The procedure for establishing mediation, i.e., determining that an independent variable $X$ affects a dependent variable $Y$ through some mediator $M$, has been under debate. The classic causal steps require that a ``total effect'' be significant, now also known as statistically acknowledged. It has been shown that the total-effect test can erroneously reject competitive mediation and is superfluous for establishing complementary mediation. Little is known about the last type, indirect-only mediation, aka ``full" or ``complete" mediation, in which the indirect ($ab$) path passes the statistical partition test while the direct-and-remainder ($d$) path fails.  This study 1) provides proof that the total-effect test can erroneously reject indirect-only mediation, including both sub-types, assuming least square estimation (LSE) $F$-test or Sobel test; 2) provides a simulation to duplicate the mathematical proofs and extend the conclusion to LAD-$Z$ test; 3) provides two real-data examples, one for each sub-type, to illustrate the mathematical conclusion; 4)  in view of the mathematical findings, proposes to revisit concepts, theories, and techniques of mediation analysis and other causal dissection analyses, and showcase a more comprehensive alternative, process-and-product analysis (PAPA).

\vspace{9pt}
\par
\end{quotation}\par

\def\thefigure{\arabic{figure}}
\def\thetable{\arabic{table}}

\renewcommand{\theequation}{\thesection.\arabic{equation}}

\fontsize{12}{14pt plus.8pt minus .6pt}\selectfont

\section{Introduction}
%the classic mediation model

The procedure to establish mediation, i.e., how an independent variable $X$ affects a dependent variable $Y$ through some mediator $M$, has been under debate. The classic \textit{causal-steps} procedure requires that the \textit{total effect}  ($c$), i.e., the effect of $X$ on $Y$ without controlling $M$, be significant, now known as statistically acknowledged \citep{baron1986}. It has been shown that the total-effect test can erroneously reject competitive mediation and is superfluous for establishing complementary mediation \citep{jiang2021total,hayes2009beyond, zhao2010reconsidering}. Little is known about the third and the last type, the indirect-only mediation, in which the indirect ($ab$) path passes the statistical test while the direct-and-remainder ($d$) path fails. 

Roughly equivalent to ``full mediation'' aka ``complete mediation" in the classic quasi-typology of \textit{full, partial, and no mediation}, indirect-only mediation is believed to be the strongest form of mediation 
\citep[p. 126]{baron1986,Hayes2022IntroMediation}. While a revised procedure of causal steps \citep{kenny1998data, kenny2008reflections, kenny2021mediation} allows researchers to ``suspend'' or ``relax'' the total-effect test when \textit{suppression}, aka \textit{inconsistent} or \textit{competitive mediation}, is suspected, full (indirect-only) mediation and partial (complementary) mediation do not qualify for the relief.  As of today, the total-effect test and, most importantly, the underlying conception of ``mediation" and ``effect" remain at the core of the criteria for establishing mediation across disciplines and languages (e.g. \cite{kenny2021mediation}; \cite{mathieu2006clarifying}; \cite{rose2004mediator}; \cite{wen2004testing}; \cite{wen2005compare}; \cite{wen2014analyses}; \cite{wen2022review}). Section 2 below provides a brief review of the debate over the total-effect test.

This study is assigned several tasks. 1a) Provide a mathematical proof that the total-effect test can erroneously reject indirect-only mediation, including both sub-types, assuming least square estimation (LSE) and $F$-test. 1b) Provide derivation to show that the same  results can be obtained assuming Sobel test. 
2) Provide a simulation to duplicate the mathematical proofs and extend the conclusion to LAD-$Z$ test. 
% and estimate the probabilities of such errors. 
3) Provide two real-data examples, one for each sub-type of the indirect-only mediation, to illustrate the mathematical proof and the simulation outcomes.  4a) In light of the mathematical findings, propose revisions to the concepts, theories, and techniques of mediation analysis and other causal dissection analyses. 4b) Introduce the principles of a more comprehensive, i.e., more encompassing and more informative, alternative,  \textit {process-and-product analysis} (PAPA).

\section{Debate on Total-Effect Test for Establishing Mediation}

Mediation model suggests a causal chain where an independent variable $X$ affects a dependent variable $Y$ through a third variable $M$, known as a \textit{mediator}.  In a classic work that influenced generations of researchers, \cite{baron1986} defines mediation as a linear regression model: 
\begin{eqnarray}
    \label{eq: model1}
    M &=& i_M + aX + \varepsilon_M,\\
    \label{eq: model2}
    Y &=& i_Y + bM + dX + \varepsilon_Y,
\end{eqnarray}
where the errors are assumed to follow independent normal distributions: 
\[\begin{split}
    \varepsilon_M \sim N(0,\sigma^2_M),\quad 
    \varepsilon_Y \sim N(0,\sigma^2_Y).
\end{split}\]

\begin{figure}[t!]
    \centering
    \includegraphics[scale=0.75]{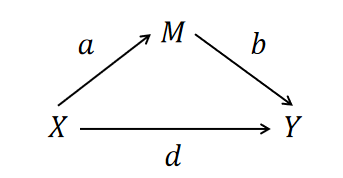}
    \caption{Mediation model proposed by \cite{baron1986}.}
    \label{fig: Model}
\end{figure}

%direct, indirect and total effect
As shown in Figure \ref{fig: Model}, the model involves two paths: 1) the \textit{indirect path} $``X \rightarrow M \rightarrow Y"$ indicates the \textit{mediated effect} of $X$ on $Y$ via \textit{mediator} $M$, which equals to $a \times b$, and 2) the so-called \textit{direct path} $``X \rightarrow Y \mid M"$ indicates the \textit{direct-and-remainder effect} of $X$ on $Y$ while $M$ is controlled, represented by $d$.
Reorganizing models \eqref{eq: model1} and \eqref{eq: model2}, we have: 
\begin{equation}\label{eq: model3}
    Y = i_Y^* + cX + \varepsilon_Y^*,
\end{equation}
where $i_Y^* = i_Y + bi_M$, $\varepsilon_Y^* = \varepsilon_Y + b\varepsilon_M$, and $c=a \times b + d$ represents the \emph{total effect} of $X$ on $Y$. 

%three types of mediation
A formal typology has been established that features three types of mediation \citep{zhao2010reconsidering,zhao2011does}: (1) \textit{Complementary mediation}, where mediated effect and direct effect both pass the statistical partition tests and bear the same sign, i.e., $a\times b\times d>0$; (2) \textit{Competitive mediation,} where mediated effect and direct effect both pass the tests and bear opposite signs, i.e., $a\times b\times d<0$; and (3)  \textit{Indirect-only mediation,} where mediated effect passes the test while direct effect fails, i.e., $a\times b\neq 0$ but $d=0$. 
%To be more specific, indirect-only mediation can be further divided into two different sub-types based on the estimated effects: directionally complementary and directionally competitive, where the estimated value $\hat{a}\times \hat{b}$ and $\hat{d}$ bear the same or different signs, i.e., $\hat{a}\times \hat{b} \times \hat{d} > 0$ or $<0$, respectively.
%[Comment to TX: can these happen? can we define them mathematically?]} 

Almost all experts accept and adopt the above definition of  mediation, namely a statistically acknowledged $a \times b$. The ``causal-steps'' procedure, however, adds another test, a statistically significant total effect, $c = a \times b + d$. This ``total-effect test" is necessary because, according to this dominant doctrine, \textit{c} represents \textit{the effect to be mediated}; a statistically non-significant \textit{c} indicates there is nothing to mediate hence no mediation is possible. Therefore, in the causal-steps procedure, if \textit{c} fails to pass the statistical test, the mediation hypothesis is declared a failure and further analysis is stopped. 

Although the causal-steps doctrine dominated mediation analysis across disciplines, whether and under what conditions the total-effect test should be required became a subject of discussion and debate. The opinions may be organized into three groups.

\textit{1) Complete acceptance}: The seminal \cite{baron1986} requires the total-effect test as the first bar to pass and allows no exception for establishing mediation of any type. The procedure and the total-effect test as the first criterion have been recommended time and again by mediation experts across disciplines (e.g., \cite{judd1981,mathieu2006clarifying,rose2004mediator}).  As we write, the total-effect test, and more importantly the underlying conception of ``mediation" and ``effect," remain part of the standard guidelines for establishing mediation across disciplines and languages (e.g. \cite{kenny2021mediation}; \cite{mathieu2006clarifying,rose2004mediator}\cite{wen2004testing,wen2005compare,wen2014analyses,wen2022review}).

\textit{2) Conditional Suspension.} Even before \cite{baron1986}, statisticians had recognized “suppression”, aka “inconsistent models”, “confounding”, and, more recently, ‘‘competition", where the mediated path $a\times b$ and the direct-and-remainder path $d$ have the opposite signs \citep{breslow1980statistical,davis1985logic,judd1981estimating,lord1968statistical,mcfatter1979use,velicer1978suppressor}.  The subject came up more often after \cite{baron1986}, as shown in numerous studies discussing it \citep{cliff1994all,cohen2013applied,conger1974revised, collins1998alternative,hamilton1987sometimes,hayes2009beyond,horst1941role,kenny1998data,kenny2021mediation,mackinnon2000equivalence,rucker2011mediation,sharpe1997relationship,shrout2002mediation,tzelgov1991suppression,zhao2010reconsidering}.
% (Breslow \& Day, 1980; Cliff \& Earleywine, 1994; Cohen \& Cohen, 1983; Conger, 1974; Collins et al., 1998; Davis, 1985; Hamilton, 1987; Hayes, 2009; Horst, 1941; Judd and Kenny, 1981b; Kenny et al., 1998; Kenny, 2003; MacKinnon, Krull, \& Lockwood, 2000; Lord \& Novick, 1968; Rucker et al, in print; Sharpe \& Roberts, 1997; McFatter, 1979; Shrout \& Bolger, 2002; Tzelgov \& Henik, 1991; Velicer, 1978; Zhao, Lynch \& Chen, 2010). 

It was 
\cite{collins1998alternative}
% Collins et al.’s (1998) 
who proposed suspending the total-effect test for inconsistent mediation. Only a special type of the ``inconsistent" models -- those with dichotomous variables -- would qualify for the relief. Other authors, at about the same time or shortly after, 
suggested suspending or dropping the total-effect test when there is \textit{a priori} belief of suppression \citep{kenny1998data,kenny2021mediation,mackinnon2000contrasts,mackinnon2000equivalence,mackinnon2002comparison,shrout2002mediation}.  
\citet[p. 438]{shrout2002mediation} also added a type, by proposing to relax the total-effect test for distal processes and expectedly small effect sizes.
% Shrout and Bolger (2002, p. 438) 

These authors often stressed the importance of retaining the total-effect test for all other types of mediation. 
\citet[p. 430]{shrout2002mediation}, for example, argued that the total-effect test has conceptual usefulness, because ``clearly, (researchers) need to first establish that the effect exists''.  

\textit{3) Complete repeal}: \citet[p. 414]{hayes2009beyond} pointed out that suppression is a regular occurrence and recommended ``researchers not require a significant total effect”.  \citet[p. 199]{zhao2010reconsidering} re-conceptualized “suppression” as “competitive mediation”, and presented a real-data example in which the competition caused a non-significant \textit{c} even though the mediated path was strong.  They hence concluded ``There need not be a significant zero-order effect of $X$ on $Y,\ldots,$ to establish mediation”. \citet[p. 18]{rucker2011mediation} agreed, concluding that ``focusing on the significance of the $X\rightarrow Y$ relationship before or after examining a mediator might be unnecessarily restrictive”.

The discussions and debates, however, were conducted mostly on conceptual and philosophical levels without mathematically rigorous evidence. In fact, the subjects under discussion were not formulated as mathematical problems until recently.

\cite{zhao2010reconsidering} and \cite{zhao2011does} replaced the traditional one-dimension conception of mediation with the two-dimension framework, which allowed the authors to replace the dominant full-partial-no quasi-typology with the five-type typology. Through the prism of the new typology, and armed with the new vernaculars, the authors observed that the total-effect test may 1) be superfluous for establishing complementary mediation 2) erroneously reject competitive mediation,  and 3) erroneously reject indirect-only mediation. 

Although not backed by rigorous proof or systematic evidence other than the one real-data example \citep{zhao2010reconsidering}, the three observations fostered three questions: Does the total-effect test help, harm, or neither for establishing each of the three types, i.e., complementary, competitive, or indirect-only mediation?

 \cite{jiang2021total} turned one observation, that the total-effect test is superfluous, into a mathematical conjecture, and produced a proof for the conjecture. They did so after transforming the task into a series of mathematical problems, which is to verify the geometry of rejection regions of hypothesis tests for $a$, $b$, $d$ and $c$.
Employing theoretical analyses, mathematical proofs, Monte Carlo simulation, and real-data examples, \cite{jiang2021total}  demonstrated that the total-effect test is indeed superfluous for establishing complementary mediation when the paths are estimated by the least square estimators and tested by $F$- or Sobel test.

We are to extend the advancement to the other two types. The geometric analysis developed by \cite{jiang2021total} potentially can be applied to build a mathematical framework for analyzing mediation of all types. The potential, however, has yet to be realized. This study is to fill the gap. Section 3 reviews the key elements of the geometric analysis, and extends the analysis to establishing complementary mediation under the LSE-Sobel frameworks. Sections 4 and 5 utilize the geometric approach to analyze indirect-only and competitive mediation under LSE-$F$ and LSE-Sobel frameworks. The results are validated numerically in Section 6 via simulation. Section 7 applies the main results to two real-data examples. Section 8 summarizes and discusses the main findings.
%adopts the same structure as  \cite{jiang2021total} to provide mathematical proof of the claim.

\section{A Geometric Perspective to Study the 
Total-Effect Test}

\subsection{Geometric Representation of Criteria for Establishing Mediation}\label{subsec:KeyIdea}
% {\red [Comment to Tingxuan: we need to establish the general framework study total-effect test via geometric analysis, and show clear logic why the debate about total-effect test can be resolved by checking the geometry of the corresponding rejection regions.]}
%For the unknown model parameters $(a,b,d,c)$, let $(\hat a,\hat b,\hat d,\hat c)$ be their estimators and $(\cR_a,\cR_b,\cR_d,\cR_c)$ be the rejection regions of hypothesis tests for testing regression coefficients $(a,b,d,c)$ in regression models \eqref{eq: model1}-\eqref{eq: model3}.

% In practice, we rely on hypothesis tests to determine direct, indirect, and total effects from observed data. The rejection regions of these hypothesis tests with significance level $\alpha$, denoted as $\cR_{a\times b}(\alpha)$, $\cR_d(\alpha)$, and $\cR_c(\alpha)$, provide a basis for claiming statistically significant causal effects. By examining the sign of their estimators $(\hat a, \hat b, \hat d, \hat c)$, we can determine the nature of the causal effect. For instance, a positive direct causal effect is claimed when $\hat d > 0$ and the observed data fall within the rejection region $\cR_d(\alpha)$. Similar principles apply to the existence and sign of total and indirect effects.

In data analysis, researchers use hypothesis tests to determine whether the direct, indirect, and total effects each pass the partition threshold \citep{kenny1998data, zhao2010reconsidering}.
% , as the parameters $(a, b, d)$ and $c$ are unknown. 
Denoting the rejection regions of appropriate hypothesis tests with significance level $\alpha$ as $\cR_{a\times b}(\alpha)$, $\cR_d(\alpha)$, and $\cR_c(\alpha)$, and using estimators $(\hat a, \hat b, \hat d, \hat c)$, we can claim a statistically significant causal effect based on the sign of its estimator. For example, a positive direct causal effect is claimed when $\hat d > 0$ and the observed data fall within
% the rejection region 
$\cR_d(\alpha)$. 
Similar definitions apply to
the 
% existence and sign of 
total and indirect effects.

The \textit{process-and-product approach} (PAPA) defines three types of mediation, namely complementary ($\bC_+$), indirect-only ($\bC_0$), and competitive ($\bC_-$), at a given significance level $\alpha$  \citep{jiang2021total, Liu2023Electronic, zhao2010reconsidering,zhao2011does}. The \textit{causal steps} approach, however, requires also a statistically significant total effect (\textit{c}) to claim any type of mediation \citep{baron1986}. The two sets of requirements together  would imply the following:
%the observed data fall into $\bD\in \cR_{a\times b}(\alpha)\cap\cR_d(\alpha)\cap\cR_c(\alpha)$ and $\hat a\times\hat b\times\hat d>0$.
\begin{eqnarray}
\label{eq:Principle4IndirectOnlyMediation}
&\bC_+:&\cD\in \cR_{a\times b}(\alpha)\cap\cR_d(\alpha)\cap\cR_c(\alpha)\ \mbox{and}\ \hat a\times\hat b\times\hat d>0;\\
&\bC_0:&\cD\in\cR_{a\times b}(\alpha)\cap\bar\cR_d(\alpha)\cap\cR_c(\alpha);\\
\label{eq:Principle4ComplementaryMediation}
\label{eq:Principle4CompetitiveMediation}
&\bC_-:&\cD\in\cR_{a\times b}(\alpha)\cap\cR_d(\alpha)\cap\cR_c(\alpha)\ \mbox{and}\ \hat a\times\hat b\times\hat d<0;\quad\ 
\end{eqnarray}
where $\cD$ represents the observed data, $\bar\cR$ stands for the complementary set of a rejection region $\cR$.
% If the total-effect test plays an essential role,
% is essential in the ``causal steps" criterion, 
If the total-effect test fails, the mediation fails to validate.
However, the PAPA analysts propose alternative criteria without considering the total effect $c$:
\begin{eqnarray}
\label{eq:AltPrinciple4IndirectOnlyMediation}
&\bC_+^*:&\cD\in \cR_{a\times b}(\alpha)\cap\cR_d(\alpha)\ \mbox{and}\ \hat a\times\hat b\times\hat d>0;\\
&\bC_0^*:&\cD\in\cR_{a\times b}(\alpha)\cap\bar\cR_d(\alpha);\\
\label{eq:AltPrinciple4ComplementaryMediation}
\label{eq:AltPrinciple4CompetitiveMediation}
&\bC_-^*:&\cD\in\cR_{a\times b}(\alpha)\cap\cR_d(\alpha)\ \mbox{and}\ \hat a\times\hat b\times\hat d<0.\quad\ 
\end{eqnarray}

Therefore, the need for the total-effect test in establishing mediation can be rationalized by examining the geometric relationships of the corresponding rejection regions.
Specifically, if we demonstrate that \begin{equation}\label{eq:Geometry4ComplementaryMediation}
\cR_{a\times b}(\alpha)\cap\cR_d(\alpha)\cap\bar\cR_c(\alpha)=0\ \mbox{whenever}\ \hat a\times\hat b\times\hat d>0,
\end{equation} we establish $\bC_+\Leftrightarrow\bC_+^*$, indicating that the total-effect test is unnecessary for complementary mediation. Similarly, if we show that \begin{equation}\label{eq:Geometry4IndirectOnlyMediation}
\cR_{a\times b}(\alpha)\cap\Bar{\cR}_d(\alpha)\cap\Bar{\cR}_c(\alpha)\neq \emptyset,
\end{equation} 
we have $\bC_0\nLeftrightarrow\bC_0^*$, suggesting that the total effect test may lead to misleading results and erroneously reject indirect-only mediation when we consider criterion $\bC^*_0$ to be more appropriate. Likewise, if we demonstrate that \begin{equation}\label{eq:Geometry4CompetitiveMediation}
\cR_{a\times b}(\alpha)\cap\cR_d (\alpha)\cap \Bar{\cR}_c(\alpha)\not= \emptyset\ \mbox{whenever}\ \hat a\times\hat b\times\hat d<0,
\end{equation} we have $\bC_-\nLeftrightarrow\bC_-^*$, implying that the total effect test may incorrectly reject competitive mediation.
% Through geometric analysis, we can address the  debate on the necessity of the total-effect test in establishing mediation.
In the following sections, we will provide a detailed implementation of the above geometric analysis.
{

\subsection{%Different Approaches for 
Estimating and Testing Mediation Effects}

%The LSE-F framework
The LSE-$F$ framework proposed by \cite{judd1981} estimates the parameters $(a, b, d, c)$ using least squares estimators (LSEs) and tests their significance using $F$-tests. The following equations define the LSEs:
\begin{eqnarray*}(\hat{i}_M,\hat{a},\hat{i}_Y,\hat{b},\hat{d})&=&\big(
\bM^T\mathcal{D}_{1,X}
\big(\mathcal{D}_{1,X}^T\mathcal{D}_{1,X}\big)^{-1},\bY^T\mathcal{D}_{1,M,X}
\big(\mathcal{D}_{1,M,X}^T\mathcal{D}_{1,M,X}\big)^{-1}\big),\\
\hat{c}&=&\hat{a}\times\hat{b}+\hat{d},
\end{eqnarray*}
where $\mathcal{D}_{1,X} = (\bm 1, \bX)$ and $\mathcal{D}_{1,M,X} = (\bm 1, \bM, \bX)$ are data matrices with or without the mediator, respectively. The rejection regions for the $F$-tests on $(a, b, d, c)$ with significance level $\alpha$ are defined as follows:
\begin{eqnarray}
    \label{eq: rejection_a}
    \cR_a(\alpha) &=& \left\{\frac{||\bm M_{\bm 1,\bX}-\bM_{\bm 1}||/(2-1)}{||\bM-\bM_{\bm 1,\bX}||/(n-2)} > \lambda_{1,n-2}(\alpha)\right\},\\
    \label{eq: rejection_b}
    \cR_b(\alpha) &=& \left\{\frac{||\bY_{\bm 1,\bM,\bX}-\bY_{\bm 1,\bX}||/(3-2)}{||\bY-\bY_{\bm 1,\bM,\bX}||/(n-3)} > \lambda_{1,n-3}(\alpha)\right\},\\
    \label{eq: rejection_d}
    \cR_d(\alpha) &=& \left\{\frac{||\bY_{\bm 1,\bM,\bX}-\bY_{\bm 1,\bM}||/(3-2)}{||\bY-\bY_{\bm 1,\bM,\bX}||/(n-3)} > \lambda_{1,n-3}(\alpha)\right\},\\
    \label{eq: rejection_c}
    \cR_c(\alpha) &=& \left\{\frac{||\bY_{\bm 1,\bX}-\bY_{\bm 1}||/(2-1)}{||\bY-\bY_{\bm 1,\bX}||/(n-2)} > \lambda_{1,n-2}(\alpha)\right\},
\end{eqnarray}
where $\bY=(Y_1,\ldots,Y_n)^T$, $\bY_{\bX}$ represents the projection of $\bY$ onto the linear space spanned by $\bX$, and $\lambda_{t,s}(\alpha)$ is the $\alpha$-quantile of $F$-distribution with degrees of freedom $(t,s)$. Additionally, to claim the significance of the indirect effect $a\times b$, we reject the hypothesis test when both $a$ and $b$ are rejected, i.e., $\cR_{a\times b}(\alpha)$ is replaced with $\cR_a(\alpha)\cap\cR_b(\alpha)$.

%The LSE-Sobel framework
Alternatively, \cite{baron1986} suggested the LSE-Sobel framework, which is similar to the LSE-$F$ framework but uses the Sobel test to examine the indirect effect $a \times b$. The test statistic $S$ is defined as:
\begin{equation}\label{eq:SobelTest_Statistics}
    S = \frac{\hat{a}\hat{b}}{\left(\hat{a}^2 \bbV(\hat{b}) + \hat{b}^2 \bbV(\hat{a})\right)^{1/2}}.
\end{equation}
Under the null hypothesis of $a\times b=0$, $S$ asymptotically follows a standard normal distribution. The rejection region for the Sobel test is given by:
\begin{equation}\label{eq:RejectionRegion_Sobel}
\cR_{a\times b}(\alpha)=\left\{|S| 
% = \left|\frac{\hat{a}\hat{b}}{\left(\hat{a}^2 Var(\hat{b}) + \hat{b}^2 Var(\hat{a})\right)^{1/2}}\right|
>z_{\alpha/2}\right\},
\end{equation}
where $z_\alpha$ represents the $\alpha$-quantile of standard normal distribution.
The LSE-Sobel framework provides a direct inference of the indirect effect $a \times b$ using a single test, but is limited by not being an exact test as the exact distribution of the test statistic $S$ depends on the values of $a$ and $b$.

%The LAD-Z framework
The LSE-based frameworks assume normal distribution for the noise terms $\varepsilon_M$ and $\varepsilon_Y$. 
In case of non-normality, 
an alternative is the LAD-$Z$ framework \citep{pollard1991}.
It uses the test statistic $z = |\check{\beta}|/sd(\check{\beta})$ compared to the standard normal distribution. 
Here, $\check{\beta}$ is the \emph{least absolute deviance} (LAD) estimator of $\beta$ for $\beta \in \{a,b,d,c\}$, and $sd(\check{\beta})$ is the estimated standard deviation of $\check{\beta}$.

\subsection{Transforming the Observed Data for Simpler %Geometric 
Representation}

The LSE estimators $(\hat{a},\hat{b},\hat{d},\hat{c})$ and the corresponding rejection regions in the LSE-$F$ and LSE-Sobel frameworks involve complex components. However, we have discovered a simpler mathematical formulation by properly transforming the original data matrix, inspired by \cite{jiang2021total}.
% Apparently, the mathematical formulas of the LSE estimators $(\hat{a},\hat{b},\hat{d},\hat{c})$ and corresponding rejection regions, i.e., $\cR_a(\alpha)\cap\cR_b(\alpha),\cR_d(\alpha),\cR_c(\alpha)$ and $\cR_{a\times b}(\alpha)$, involve complicated components and are in general not friendly to work with directly under either the LSE-$F$ framework or the LSE-Sobel framework.
% Fortunately, however, inspired by \cite{jiang2021total}, we found that a much simpler mathematical formulation of these elements can be derived after the original data matrix is properly transformed.

Lemma 1 in \cite{jiang2021total} demonstrated that the LSE estimators $(\hat{a},\hat{b},\hat{d},\hat{c})$ and the rejection regions for $F$-tests are invariant to scale and orthogonal transformations of the observed data. 
The following lemma extends the original lemma in \cite{jiang2021total} by including the invariance of the Sobel test for $a\times b$, and its proof can be found in Section 1 (S1) of the Supplementary Material.

% \begin{lemma}\label{lem:InvarianceUnderOrthogonalTransformation}
% For regression model 
% \[\begin{split}
%     Y = \beta_0 X_0 + \beta_1 X_1 + \cdots + \beta_p X_p + \varepsilon
% \end{split}\]
% with $\bbeta = (\beta_0,\beta_1,\ldots,\beta_p)$ and $\mathcal{D} = (\bX_0,\bX_1,\ldots,\bX_p,\bY)$
% be the coefficient vector and data matrix, respectively,
% denote $\hat{\bbeta}$ as the LSE of $\bbeta$.
% Define the transformed data matrix under scale and orthogonal transformation as $\Tilde{\mathcal{D}} = (\Tilde{\bX}_0,\Tilde{\bX}_1,\ldots,\Tilde{\bX}_p,\Tilde{\bY}) = \gamma\Gamma\mathcal{D}$ for any $\gamma > 0$ and $n\times n$ orthogonal matrix $\Gamma$, and the corresponding regression problem as 
% \[\begin{split}
%     \Tilde{Y} = \beta_0 \Tilde{X}_0 + \beta_1 \Tilde{X}_1 + \cdots + \beta_p \Tilde{X}_p + \varepsilon.
% \end{split}\]
% Let $\Tilde{\bbeta}$ be the LSE of $\bbeta$, $\Tilde{\mathcal{R}}_j (\alpha)$ be the rejection region of $F$-test for hypothesis $\beta_j = 0$, and $\Tilde{\mathcal{R}}_{i,j} (\alpha)$ be the rejection region of Sobel test for hypothesis $\beta_i \times \beta_j = 0$
% under the transformed problem. Then for any $i\not= j\in \{0,1,\ldots,p\}$ and $\alpha \in (0,1)$, we have 
% \[\begin{split}
%     \Tilde{\bbeta} = \hat{\bbeta},\ \Tilde{\mathcal{R}}_j (\alpha)= \mathcal{R}_j(\alpha),\text{ and } \Tilde{\mathcal{R}}_{i,j}(\alpha)= \mathcal{R}_{i,j}(\alpha).
% \end{split}\]

% % {\red [Comment to Tingxuan: we may need to formally define the invariance as in Yingkai's paper.]}
% \end{lemma}

\begin{lemma}\label{lem:InvarianceUnderOrthogonalTransformation}
Consider the regression model in Eq.~\eqref{eq: model1}-\eqref{eq: model3} with data matrix $\mathcal{D} = (\bm 1,\bm M, \bm X, \bm Y)$. Let $\Tilde{\mathcal{D}} = \gamma\Gamma\mathcal{D}$ be the transformed data matrix under scale and orthogonal transformations, where $\gamma > 0$, $\Gamma$ orthogonal matrix. In the transformed problem, the LSE $\Tilde{\beta}$ and rejection regions $\Tilde{\mathcal{R}}_{\beta} (\alpha)$ for the $F$ and Sobel test ($\beta\in{a,b,d,c,a\times b}$)
% , and the rejection region $\Tilde{\mathcal{R}}_{a\times b} (\alpha)$ for the Sobel test 
remain the same as the original problem. 
% For regression model defined in Eq.~\eqref{eq: model1}-\eqref{eq: model3} with data matrix $\mathcal{D} = (\bm 1,\bm M, \bm X, \bm Y)$, 
% define the transformed data matrix under scale and orthogonal transformation as $\Tilde{\mathcal{D}} = (\Tilde{\bm 1}, \Tilde{\bM},\Tilde{\bX},\Tilde{\bY}) = \gamma\Gamma\mathcal{D}$ for any $\gamma > 0$.
% and $n\times n$ orthogonal matrix $\Gamma$, and the corresponding regression problem as 
% \begin{eqnarray}
%     \label{eq: transformedmodel1}
%     \Tilde{M} &=& i_M\Tilde{1} + a\Tilde{X} + \varepsilon_M,\\
%     \label{eq: transformedmodel2}
%     \Tilde{Y} &=& i_Y\Tilde{1} + b\Tilde{M} + d\Tilde{X} + \varepsilon_Y.
% \end{eqnarray}
% Let $(\Tilde{a},\Tilde{b},\Tilde{d},\Tilde{c})$ be the LSE of $(a,b,d,c)$, $\Tilde{\mathcal{R}}_{\beta} (\alpha)$ be the rejection region of $F$-test for $\beta\in\{a,b,d,c\}$, and $\Tilde{\mathcal{R}}_{a\times b} (\alpha)$ be the rejection region of Sobel test for hypothesis $a \times b = 0$
% under the transformed problem. Then for any $\beta\in \{a,b,d,c\}$ and $\alpha \in (0,1)$, we have 
% \[\begin{split}
%     \Tilde{\beta} = \hat{\beta},\ \Tilde{\mathcal{R}}_{\beta} (\alpha)= \mathcal{R}_{\beta}(\alpha),\text{ and } \Tilde{\mathcal{R}}_{a\times b}(\alpha)= \mathcal{R}_{a\times b}(\alpha).
% \end{split}\]
% {\red [Comment to Tingxuan: we may need to formally define the invariance as in Yingkai's paper.]}
\end{lemma}
%\begin{lemma}\label{lem:OrthogonalTransformationOfData}

The above lemma suggests that we can transform the original data matrix to obtain simpler rejection regions.
As highlighted by \cite{jiang2021total}, for a classic mediation model with the data matrix 
$\mathcal{D}$ with rank$(\mathcal{D}) = 4$, there always exists an $n \times n$ real orthogonal matrix $\Gamma$ and a global scale parameter $\gamma > 0$ s.t. the transformed data matrix $\Tilde{\mathcal{D}}= (\Tilde{\bm 1}, \Tilde{\bM},\Tilde{\bX},\Tilde{\bY}) $ satisfies    $\Tilde{\bm 1} = (1,0,\ldots,0)^T$, $\Tilde{\bm X} = (x_1,x_2,0,\ldots,0)^T$, $\Tilde{\bm M} = (m_1,m_2,m_3,0,\ldots,0)^T$, $\Tilde{\bm Y} = (y_1,y_2,y_3,y_4,0,\ldots,0)^T$ with $x_2 > 0$, $m_3 > 0$ and $y_4>0$. 
% The above lemma implies that we can transform the original data matrix for simpler form of rejection regions of interest.
% As pointed out by \cite{jiang2021total}, as long as the data matrix of the classic mediation model  $\mathcal{D} = (\bm 1, \bm X, \bm M, \bm Y)$ satisfies that rank$(\mathcal{D}) = 4$, we can always find an $n \times n$ real orthogonal matrix $\Gamma$ and a global scale parameter $\gamma > 0$ s.t. the transformed data matrix $\Tilde{\mathcal{D}}= \gamma\Gamma \mathcal{D} =(\Tilde{\bm 1},\Tilde{\bm X},\Tilde{\bm M},\Tilde{\bm Y})$ satisfying    $\Tilde{\bm 1} = (1,0,\ldots,0)^T$, $\Tilde{\bm X} = (x_1,x_2,0,\ldots,0)^T$, $\Tilde{\bm M} = (m_1,m_2,m_3,0,\ldots,0)^T$, $\Tilde{\bm Y} = (y_1,y_2,y_3,y_4,0,\ldots,0)^T$ with $x_2 > 0$, $m_3 > 0$ and $y_4>0$. 
% %\end{lemma}

Apparently, the transformed data $\Tilde{\cD}$ simplifies the LSE estimators and rejection regions. The following lemma summarizes the results, including the explicit form of $\cR_{a\times b}$, which was not provided in \cite{jiang2021total}.
% such a $\Tilde{\cD}$ would lead to much simpler expressions of the LSE estimators and rejection regions of interest.
% Lemma \ref{lem:LSE-F|Sobel} below summarizes the results accordingly, with the explicit form of $\cR_{a\times b}$ detailed, which is absent in \cite{jiang2021total}.
%for rejection region of the Sobel test, i.e., $\cR_{a\times b}(\alpha)$.

% Compared to the original data matrix $\cD$, the transformed data matrix $\Tilde{\cD}$ enjoys a concise form and would lead to much simpler expressions of LSE estimators and the corresponding rejection regions under $F$ test or Sobel test according to Lemma~\ref{lem:InvarianceUnderOrthogonalTransformation}.
% The Lemma \ref{lem:LSE-F|Sobel} below summarizes these results, which enhances the Lemma 2 of \cite{jiang2021total} with the result for rejection region of the Sobel test, i.e., $\cR_{a\times b}$.

%Therefore, after transformation, $(\hat{a},\hat{b},\hat{d},\hat{c})$ and $(\cR_a(\alpha),\cR_b(\alpha),\cR_d(\alpha),\cR_c(\alpha))$ can have a simpler form as stated in the following lemma.

\begin{lemma}\label{lem:LSE-F|Sobel}
    %Let $\bm 1 = (1,\ldots,1)^T$ be the $n$-dimensional column vector whose elements all equal to $1$, and let $\bm X$, $\bm M$ and $\bm Y$ be the column data vectors for variables $X$, $M$ and $Y$ respectively. 
        If rank$(\Tilde{\mathcal{D}})=4$, we obtain simple expressions for the LSE estimators and rejection regions of the causal effects of interest:
    \begin{equation}\label{eq:LSE-TransformedData}
        \hat{a} = \frac{m_2}{x_2},\quad \hat{b} = \frac{y_3}{m_3},\quad \hat{c} = \frac{y_2}{x_2},\quad \hat{d} = \frac{m_3y_2 - m_2y_3}{x_2m_3};
    \end{equation}
    % the following rejection regions for the $F$-test for $(a,b,c,d)$ and the Sobel test for $a\times b$:
    \begin{eqnarray}\label{eq:RejectionRegionOfF-TransformedData}
    \mathcal{R}_a(\alpha) &=& \{r>r_{n,\alpha}\},\\
    \mathcal{R}_b(\alpha) &=& \{p>p_{n,\alpha}\},\\
    \mathcal{R}_c(\alpha) &=& \left\{q>r_{n,\alpha}(p^2+1)^{1/2}\right\},\\
    \mathcal{R}_d(\alpha) &=& 
    \begin{cases}
        \left\{|q-rp|>p_{n,\alpha}(r^2+1)^{1/2}\right\} & \text{if }\hat{a}\hat{b}\hat{c} \geq 0,\\
        \left\{|q+rp|>p_{n,\alpha}(r^2+1)^{1/2}\right\} & \text{if }\hat{a}\hat{b}\hat{c} < 0,
    \end{cases}\\
    \mathcal{R}_{a\times b}(\alpha) &=& \left\{\frac{1}{(n-2)r^2} + \frac{1}{(n-3)p^2} < \frac{1}{z_{\alpha/2}^2}\right\};
    \end{eqnarray}
where $r = |m_2|/m_3$, $p = |y_3|/y_4$, $q = |y_2|/y_4$, $r_{n,\alpha} = \left[\lambda_{1,n-2}(\alpha)/(n-2)\right]^{1/2}$,  and $p_{n,\alpha} = \left[\lambda_{1,n-3}(\alpha)/(n-3)\right]^{1/2}$, with $\lambda_{t,s}(\alpha)$ and $z_{\alpha/2}$ defined previously.
\end{lemma}
% }

% Lemma \ref{lem:LSE-F|Sobel} indicates that each of the five rejection regions of interest corresponds to a subspace in the $3$-dimensional space indexed by $(p,q,r)$,
%, which degenerates to a region in $p\operatorname{-}q$ plane $\mathcal{P}_r$ for each specific $r$.
%The simplified formulas of $(\hat{a},\hat{b},\hat{d},\hat{c})$ and $(\cR_a(\alpha),\cR_b(\alpha),\cR_d(\alpha),\cR_c(\alpha))$ showed in Lemma~\ref{lem: lemma1} 
% making it much easier to implement the geometric analysis proposed in section~\ref{subsec:KeyIdea}.

\subsection{%Geometric Analysis Reveals that 
Total-Effect Test is Superfluous for Complementary Mediation}

Using the simplified formulas in Lemma \ref{lem:LSE-F|Sobel}, \cite{jiang2021total} showed that $\cR_{a}(\alpha)\cap\cR_{b}(\alpha)\cap\cR_d(\alpha)\subseteq \cR_c(\alpha)\ \mbox{whenever}\ \hat a\times\hat b\times\hat d>0$ under mild conditions. This implies that total-effect test is superfluous for establishing complementary mediation under the LSE-$F$ framework.
Additionally, they showed that the total-effect test is also unnecessary asymptotically under the LSE-Sobel framework.
However, their analysis within the LSE-Sobel framework lacks a geometric perspective, and can be enhanced by the following theorem,
with the proof detailed in Section 1 (S1) of the Supplementary Material.

\begin{theorem}\label{thm:SobelComplementary}
    Suppose that we rely on the LSE-Sobel framework to establish mediation. 
    If rank$(\mathcal{D})=4$, as sample size $n\rightarrow\infty$, for all $\alpha \in (0,1)$, 
    \[\begin{split}
        P\left(\cD \in \mathcal{R}_{a\times b}(\alpha) \cap \mathcal{R}_{d}(\alpha) \cap \Bar{\mathcal{R}}_{c}(\alpha)\right) \rightarrow 0\ \text{for }\hat{a}\times \hat{b}\times\hat{d} >0.
    \end{split}\]
\end{theorem}

Theorem \ref{thm:SobelComplementary} implies that as the sample size $n \rightarrow \infty$,
$ \mathcal{R}_{a\times b}(\alpha) \cap \mathcal{R}_{d}(\alpha)\subseteq  {\mathcal{R}}_{c}(\alpha)$ holds asymptotically. This provides an alternative perspective supporting the argument that the total-effect test is superfluous for testing complementary mediation under the LSE-Sobel framework.

\section{Total-Effect Test can Erroneously Reject Indirect-Only Mediation}
\cite{jiang2021total} mentioned that the total-effect test can be misleading when testing indirect-only mediation under the LSE-$F$ framework. However, no technical proof was provided to support the observation, and the observation does not cover the LSE-Sobel framework. In this section, we will fill these gaps through explicit theoretical analysis that shows the potentially misleading nature of the total-effect test for establishing indirect-only mediation.

% However, their conclusion was based on graphical intuition without a detailed technical proof.
% Moreover, the situation under the LSE-Sobel framework remains an open problem. 
% In this section, we will address these gaps by showing that total-effect test can be misleading for establishing indirect-only mediation under both LSE-$F$ and LSE-Sobel frameworks with explicit theoretical analysis. 

To show that the total-effect test may 
% lead to misleading results and 
erroneously reject indirect-only mediation under the LSE-$F$ framework, we need to verify condition \eqref{eq:Geometry4IndirectOnlyMediation}: $$\mathcal{R}_a(\alpha) \cap \mathcal{R}_b(\alpha) \cap \Bar{\cR}_d(\alpha) \cap \Bar{\cR}_c(\alpha) \not= \emptyset,\ \text{for all } \alpha \in (0,1).$$ This condition can be equivalently expressed as: \begin{equation}\label{eq: equivalent}    \mathcal{R}_a(\alpha\mid r) \cap \mathcal{R}_b(\alpha\mid r) \cap \Bar{\cR}_d(\alpha\mid r) \cap \Bar{\cR}_c(\alpha\mid r) \not= \emptyset \text{ for some }r>0,\end{equation} where $\cR_\beta(\alpha|r)$ represents the intersection of $\mathcal{R}_{\beta}(\alpha)$ and the $p$-$q$ plane $\mathcal{P}_r$ for $\beta \in \{a,b,d,c\}$. Since $\mathcal{R}_a(\alpha\mid r) = \mathcal{P}_r \cap \{r>r_{n,\alpha}\} = \emptyset$ for $0 < r \leq r_{n,\alpha}$, 
% argument \eqref{eq: equivalent} is equivalent to \begin{equation}\label{eq: equivalent2}    \mathcal{R}_a(\alpha\mid r) \cap \mathcal{R}_b(\alpha\mid r) \cap \Bar{\cR}_d(\alpha\mid r) \cap \Bar{\cR}_c(\alpha\mid r) \not= \emptyset \text{ for some }r>r_{n,\alpha}.\end{equation} 
we will focus on $r > r_{n,\alpha}$. We verify the argument by considering two sub-types of indirect-only mediation separately: the case where $\hat{a}\hat{b}\hat{d}> 0$, representing \textit{directionally complementary mediation},
and the case where $\hat{a}\hat{b}\hat{d} < 0$, representing \textit{directionally competitive mediation}.

In the case where $\hat{a}\hat{b}\hat{d} > 0$, we can observe the following relationships. Corollary 1 in \cite{jiang2021total} implies that $\hat{a}\hat{b}\hat{c} > 0$ and $q > rp$. Thus, we have $\Bar{\cR}_d(\alpha\mid r)=\left\{rp < q \leq rp+ p_{n,\alpha}(r^2+1)^{1/2}\right\}$. Additionally, $\mathcal{R}_b(\alpha\mid r) = \{p>p_{n,\alpha}\}$, and  $\Bar{\cR}_c(\alpha\mid r)=\left\{0 \leq q \leq r_{n,\alpha}(p^2+1)^{1/2},p \geq 0\right\}$.
The geometry of these regions is depicted in Figure \ref{fig: Plot1}. According to Theorem 1 in \cite{jiang2021total}, we have $r_{n,\alpha}(p^2+1)^{1/2} < rp + p_{n,\alpha}(r^2+1)^{1/2}$ for $r > r_{n,\alpha}$. Therefore, the intersection $\mathcal{R}_a(\alpha\mid r) \cap \mathcal{R}_b(\alpha\mid r) \cap \Bar{\cR}_d(\alpha\mid r) \cap \Bar{\cR}_c(\alpha\mid r)$ is  $$\left\{rp < q \leq r_{n,\alpha}(p^2+1)^{1/2},p_{n,\alpha} < p < r_{n,\alpha}/(r^2-r_{n,\alpha}^2)^{1/2}\right\},$$ which can be verified to be not empty for $r_{n,\alpha} < r < r_{n,\alpha}(1+1/p_{n,\alpha}^2)^{1/2}$. Figure \ref{fig: Plot1} (D) provides a graphical demonstration of this intersection.

% Therefore, we can see that $\mathcal{R}_a(\alpha\mid r) \cap \mathcal{R}_b(\alpha\mid r) \cap \Bar{\cR}_d(\alpha\mid r) \cap \Bar{\cR}_c(\alpha\mid r) \not= \emptyset$ if and only if $r_{n,\alpha}/
% (r^2-r_{n,\alpha}^2)^{1/2} > p_{n,\alpha}$, which is equivalent to $r < r_{n,\alpha}(1+1/p_{n,\alpha}^2)^{1/2}$. This implies that the total-effect test can reject directionally complementary indirect-only mediation erroneously when $r_{n,\alpha} < r < r_{n,\alpha}\sqrt{1+1/p_{n,\alpha}^2}$. 
% Figure \ref{fig: Plot1} provides a graphical demonstration of the geometry of $\mathcal{R}_b(\alpha\mid r)$, $\Bar{\cR}_c(\alpha\mid r)$, $\Bar{\cR}_d(\alpha\mid r)$ and their intersection under directionally complementary scenario when $r_{n,\alpha} < r < r_{n,\alpha}\sqrt{1+1/p_{n,\alpha}^2}$.

\begin{figure}[t!]
    \centering
    \includegraphics[scale=0.07]{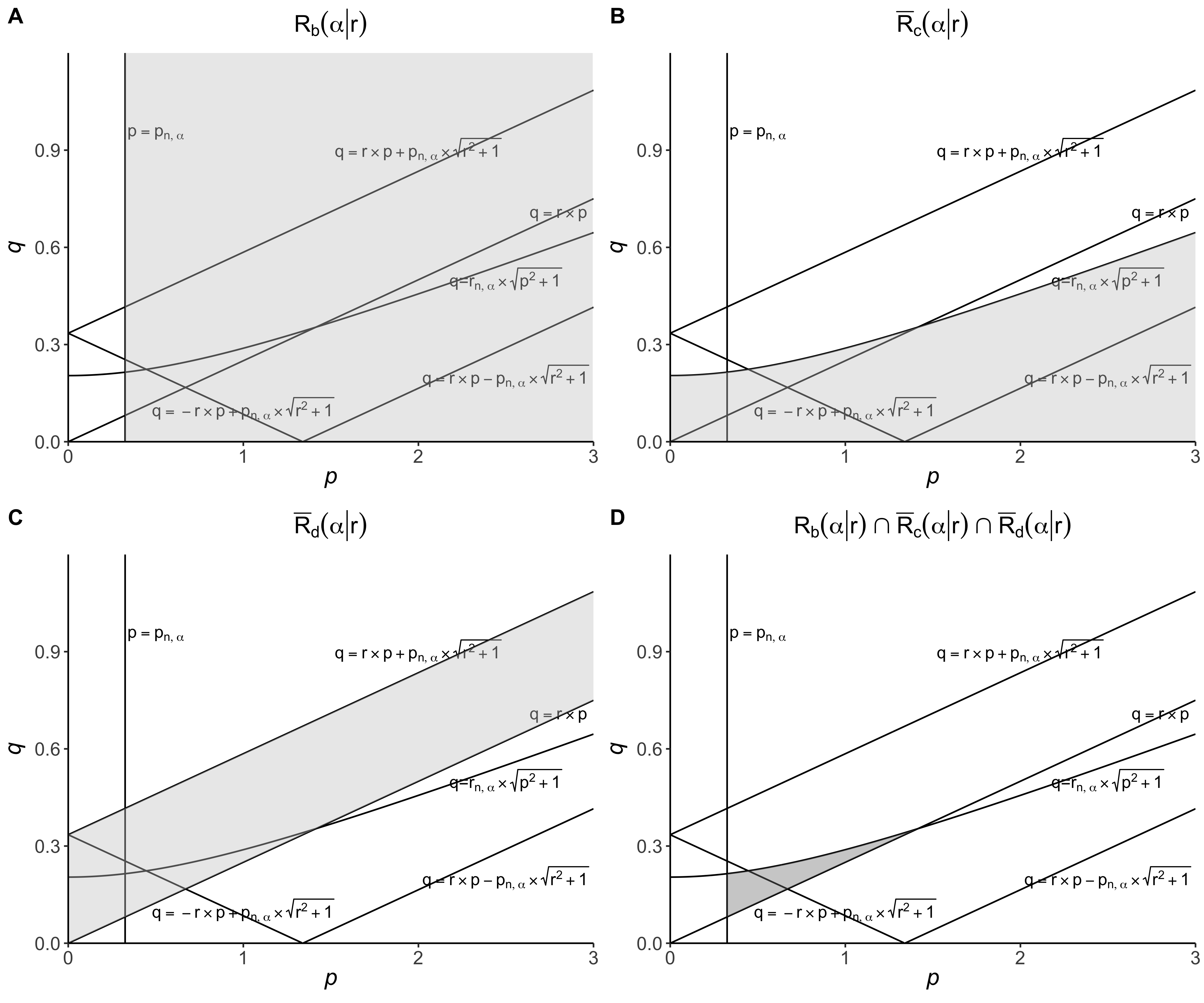}
    \caption{A graphical illustration of $\mathcal{R}_b(\alpha\mid r)$, $\Bar{\cR}_c(\alpha\mid r)$, $\Bar{\cR}_d(\alpha\mid r)$ and their intersection in  $p\operatorname{-}q$ plane for a fixed $r_{n,\alpha} < r < r_{n,\alpha}(1+1/p_{n,\alpha}^2)^{1/2}$ for directionally complementary indirect-only mediation, i.e., $\hat{a}\hat{b}\hat{d} > 0$.}
    \label{fig: Plot1}
\end{figure}

In the case where $\hat{a}\hat{b}\hat{d} < 0$, the sign of $\hat{a}\hat{b}\hat{c}$ is indeterminate. The regions $\mathcal{R}_a(\alpha\mid r)$, $\mathcal{R}_b(\alpha\mid r)$, and $\Bar{\cR}_c(\alpha\mid r)$ remain the same as in the case where $\hat{a}\hat{b}\hat{d} > 0$. The only difference lies in the region $\Bar{\cR}_d(\alpha\mid r)$. The following lemma helps define ${\cR}_d(\alpha\mid r)$ when    $\hat{a}\hat{b}\hat{c} \geq 0$.

\begin{lemma}\label{lem:competitive}
If $\hat{a}\hat{b}\hat{d} < 0$ and $\hat{a}\hat{b}\hat{c} \geq 0$, we have $q<rp$, and thus, $\mathcal{R}_d(\alpha) = \{q < rp - p_{n,\alpha}(r^2+1)^{1/2},p\geq 0\}$.
\end{lemma}
Using Lemma \ref{lem:competitive}, It can be verified that 
\[\begin{split}
    \Bar{\cR}_d(\alpha\mid r) = \begin{cases}
        \left\{rp- p_{n,\alpha}(r^2+1)^{1/2} \leq q < rp,p\geq 0\right\}, & \hat{a}\hat{b}\hat{c} \geq 0,\\
        \left\{0 \leq q \leq -rp+ p_{n,\alpha}(r^2+1)^{1/2}, p \geq 0\right\}, & \hat{a}\hat{b}\hat{c} < 0,
    \end{cases}
\end{split}\]
and the intersection of $\mathcal{R}_b(\alpha\mid r)$, $\Bar{\cR}_c(\alpha\mid r)$ and $\Bar{\cR}_d(\alpha\mid r)$ is not empty for any $r>r_{n,\alpha}$. The geometry of $\Bar{\cR}_d(\alpha\mid r)$ and the intersection of interest
% $\mathcal{R}_a(\alpha\mid r) \cap \mathcal{R}_b(\alpha\mid r) \cap \Bar{\cR}_d(\alpha\mid r) \cap \Bar{\cR}_c(\alpha\mid r)$
under the directionally competitive mediation are shown in Figure \ref{fig: Plot2}.

\begin{figure}[t!]
    \centering
    \includegraphics[scale=0.07]{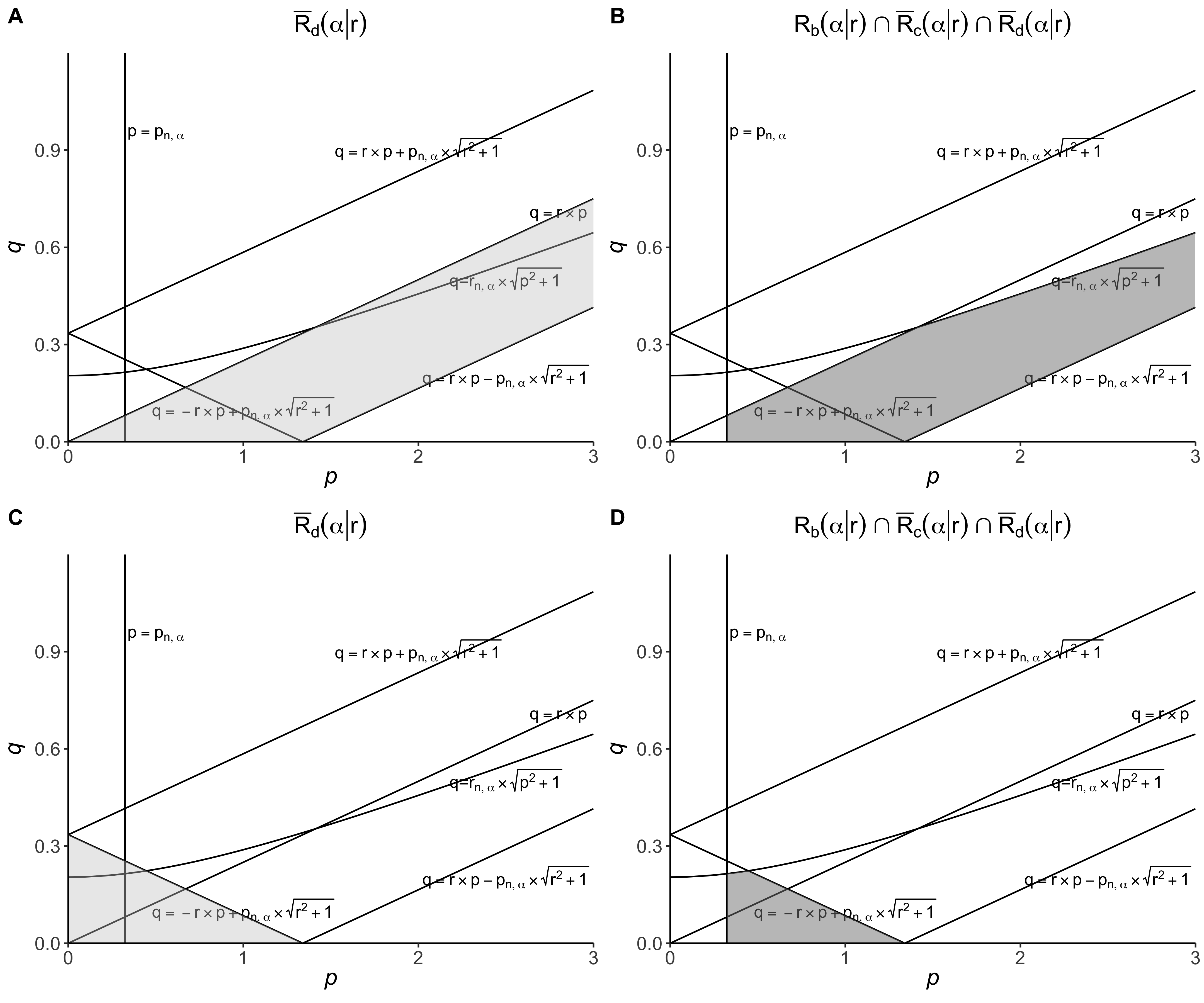}
    \caption{Geometry of $\Bar{\cR}_d(\alpha\mid r)$ and $\mathcal{R}_a(\alpha\mid r) \cap \mathcal{R}_b(\alpha\mid r) \cap \Bar{\cR}_d(\alpha\mid r) \cap \Bar{\cR}_c(\alpha\mid r)$ with fixed $r > r_{n,\alpha}$ for directionally competitive indirect-only mediation, i.e., $\hat{a}\hat{b}\hat{d} < 0$, when (A)(B) $\hat{a}\hat{b}\hat{c} \geq 0$ and (C)(D) $\hat{a}\hat{b}\hat{c} < 0$.}
    \label{fig: Plot2}
\end{figure}

Above all, we validate the argument \eqref{eq: equivalent}.
Figures \ref{fig: Plot1} and \ref{fig: Plot2} imply that $\mathcal{R}_a(\alpha\mid r) \cap \mathcal{R}_b(\alpha\mid r) \cap \Bar{\cR}_d(\alpha\mid r) \cap \Bar{\cR}_c(\alpha\mid r)$ has larger support under directionally competitive mediation, indicating a higher probability of observing an insignificant total effect when $\hat{a}\hat{b}$ and $\hat{d}$ have opposite signs and $\hat{d}$ is not statistically significant.
% This is consistent with the intuition that when $\hat{a}\hat{b}$ and $\hat{d}$ bear opposite signs and $\hat{d}$ is not statistically significant, the variance induced by $\hat{d}$ may offset that of $\hat{a}\hat{b}$, leading to a higher probability of observing an insignificant total-effect. 

{
The following theorem summarizes the above analysis and depicts the rejection regions based on a nice geometry under a mild condition. 

\begin{theorem}\label{thm: theorem1}
%Suppose there are $n$ data points in the classic mediation model as in Eq.~\eqref{eq: model1}, \eqref{eq: model2} and \eqref{eq: model3}. Let $\bm 1 = (1,\ldots,1)^T$ be the $n$-dimensional column vector whose elements all equal to $1$, and let $\bm X$, $\bm M$ and $\bm Y$ be the column data vectors for variables $X$, $M$ and $Y$ respectively. Let $\mathcal{D} = (\bm 1, \bm X, \bm M, \bm Y)$ denote the data matrix of regression. 
Suppose that we rely on the LSE-$F$ framework to establish mediation. 
If rank$(\mathcal{D})=4$, we have 
$$\mathcal{R}_a(\alpha) \cap \mathcal{R}_b(\alpha) \cap \Bar{\cR}_d(\alpha) \cap \Bar{\cR}_c(\alpha) \not= \emptyset,\ \text{for all } \alpha \in (0,1).$$ 
\end{theorem}
}

Similarly, the following Theorem shows that the total-effect test can be erroneous for establishing indirect-only mediation under the LSE-Sobel framework { with large sample size}. 
The details of proof can be found in {Section S1 of the Supplementary Material}.

\begin{theorem}\label{thm:SobelIO}
    Suppose that we rely on the LSE-Sobel framework to establish mediation.
    If rank$(\mathcal{D})=4$, there exists $N>0$ such that when sample size $n>N$,
    we have 
    \[\begin{split}
 \mathcal{R}_{a\times b}(\alpha) \cap \Bar{\mathcal{R}}_{d}(\alpha) \cap \Bar{\mathcal{R}}_{c}(\alpha)\not= \emptyset,\ \text{for all }\alpha\in(0,1).
    \end{split}\]
\end{theorem}

\section{Total-Effect Test Can Erroneously Reject Competitive Mediation}
Similar analyses can be applied to competitive mediation. The following theorems demonstrate that the total-effect test can lead to erroneously rejection of competitive mediation under LSE-$F$ and LSE-Sobel frameworks, respectively, for statistical partitioning. While previous studies have shown the possibility of erroneous rejection using bootstrap tests \citep{mackinnon2000equivalence, mackinnon2007mediation, hayes2009beyond, zhao2010reconsidering} through derivations and examples, the theorems below provide rigorous proofs assuming LSE-$F$ and LSE-Sobel tests. 
% \red Equally importantly, the theorems show the precise ranges within which the erroneous rejections occur.

\begin{theorem}\label{thm: competitive}
Suppose that we rely on the LSE-$F$ framework to establish mediation. 
If rank$(\mathcal{D})=4$, 
%Suppose there are $n$ data points in the classic mediation model as in Eq.~\eqref{eq: model1}, \eqref{eq: model2} and \eqref{eq: model3}. Let $\mathcal{D} = (\bm 1, \bm X, \bm M, \bm Y)$ denote the data matrix of regression. If rank$(\mathcal{D})=4$, using LSE-$F$ test, 
condition $\hat{a}\times \hat{b}\times \hat{d} < 0$ implies
$$\mathcal{R}_a(\alpha) \cap \mathcal{R}_b(\alpha) \cap \mathcal{R}_d(\alpha) \cap \Bar{\cR}_c(\alpha) \not= \emptyset,\  \text{for all }\alpha \in (0,1).$$ 
\end{theorem}

\begin{theorem}\label{thm:SobelCompetitive}
    Suppose that we rely on the LSE-Sobel framework to establish mediation.
    If rank$(\mathcal{D})=4$, 
    % as sample size $n\rightarrow\infty$, 
    condition $\hat{a}\times \hat{b}\times \hat{d} < 0$ implies
    $$\mathcal{R}_{a\times b}(\alpha) \cap {\mathcal{R}}_{d}(\alpha) \cap \Bar{\mathcal{R}}_{c}(\alpha)\not= \emptyset,\ \text{for all }\alpha \in (0,1).$$
\end{theorem}

The procedure for proving the indirect-only mediation was applied to proving the two theorems above, of which the details are documented in Section 1 (S1) of the Supplementary Material.

\section{Simulations}
\cite{jiang2021total} conducted simulations to show that the total-effect test is unnecessary for establishing complementary mediation under LSE-$F$ and LSE-Sobel frameworks and can be misleading with the LAD-$Z$ test. The study focused on indirect-only and competitive mediation, presenting the results for indirect-only mediation in the main text and the results for competitive mediation in Section 2 (S2) of the Supplementary Material.

% \cite{jiang2021total} used simulation to demonstrate that the total-effect test is superfluous for establishing complementary mediation when employing LSE-$F$ or LSE-Sobel test, and can mislead when conducting the LAD-$Z$ test. 
% Thus the simulation experiments in this study focus on indirect-only and competitive mediation. 
% The results concerning indirect-only mediation are shown in the main text, while those concerning competitive mediation are detailed in the
% Supplementary Material. 
 
\subsection{Numerical validation of Theorem \ref{thm: theorem1}}
To validate Theorem \ref{thm: theorem1}, we generate the simulated data from model \eqref{eq: model1} and \eqref{eq: model2} as follows:
\[\begin{split}
    & n \sim \text{Unif}(\{10,\ldots,100\}),\quad (i_M,i_Y,a,b,d) \sim \text{Unif}[-1,1]^5,\\
    & X \sim N(0,1),\quad \sigma_M^2\text{ and } \sigma_Y^2\sim \text{Inv-Gamma}(1,1).
\end{split}\]
A total of $10,000$ independent datasets of different sample sizes were simulated. For each dataset, we calculated the LSEs $(\hat{a},\hat{b},\hat{c},\hat{d})$ and $p$-values $(p_a,p_b,p_c,p_d)$ under the LSE-$F$ framework. We checked if, for any fixed $\alpha \in (0,1)$, $\max(p_a,p_b) < \alpha$ and $p_d \geq \alpha$ imply $\{p_c \geq \alpha\} \not= \emptyset$.

\begin{figure}[t!]
    \centering
    \includegraphics[scale=0.55]{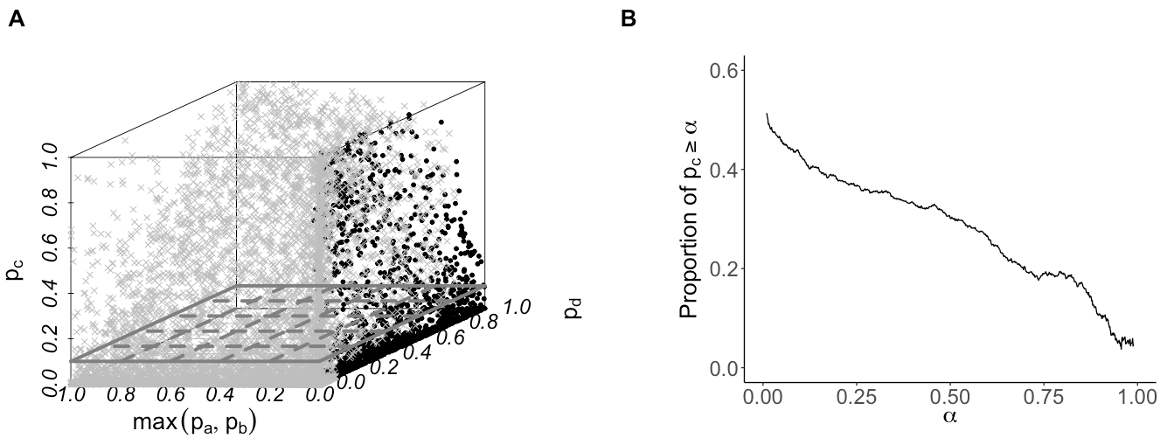}
    \caption{Numerical exploration under LSE-$F$ framework: (A) 
    % Scatter plot of 
    % under LSE-$F$ framework: 
    black solid circles for $\max(p_a,p_b) < 0.1$ and $p_d \geq 0.1$, grey crossings for $\max(p_a,p_b) \geq 0.1$ or $p_d < 0.1$, 
    and dark gray dashed plane for $p_c = 0.1$;
    (B) Proportion of datasets satisfying $p_c \geq \alpha$ when $\max(p_a,p_b) < \alpha$ and $p_d \geq \alpha$.
    % under the LSE-$F$ framework.
    }
    \label{fig: LSE-F1}
\end{figure}

\begin{figure}[t!]
    \centering
    \includegraphics[scale=0.25]{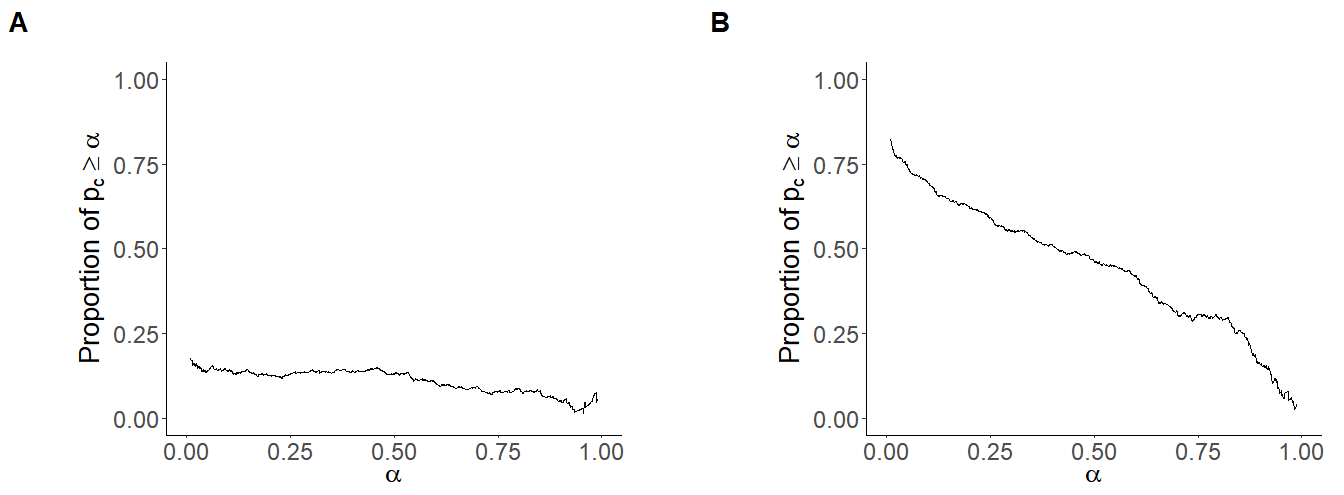}
    \caption{Proportion of datasets satisfying $p_c \geq \alpha$ when $\max(p_a,p_b) < \alpha$, $p_d \geq \alpha$ and (A) $\hat{a}\hat{b}\hat{d} > 0$, (B) $\hat{a}\hat{b}\hat{d} < 0$ under LSE-$F$ framework.}
    \label{fig: LSE-F2}
\end{figure}

Figure \ref{fig: LSE-F1} (A) checks the $p$-value condition when $\alpha=0.1$. Each simulated dataset is represented by a point in a 3-dimensional space with $\max(p_a,p_b)$, $p_d$, and $p_c$ as the $X$, $Y$, and $Z$ axes, respectively. The solid circles represents datasets satisfying $\max(p_a,p_b) < \alpha$ and $p_d \geq \alpha$, gray crossings represent data sets with $\max(p_a,p_b) \geq \alpha$ or $p_d < \alpha$, and the dark gray dashed plane represents $p_c = \alpha$. The solid circles above the plane $p_c = \alpha$ indicate the empirical set $\{p_c \geq \alpha\}$. We observe that when $\max(p_a,p_b) < \alpha = 0.1$ and $p_d \geq \alpha$, the set $\{p_c \geq \alpha\}$ is not empty. 

Figure \ref{fig: LSE-F1} (B) presents the proportion of datasets satisfying $p_c \geq \alpha$ for 1000 evenly spaced values of $\alpha$ in the range of $(0.01,0.99)$ when $\max(p_a,p_b) < \alpha$ and $p_d \geq \alpha$ under the LSE-$F$ framework. It shows that for significance levels $\alpha$ smaller than 0.1, which is commonly used in practice, the proportion of cases where $p_c > \alpha$ is greater than 40\%. This indicates a high probability of erroneous rejection of indirect-only mediation by 
the total-effect test.
% erroneously rejecting the presence of indirect-only mediation.

% To check the theoretical result for different values of $\alpha$, the proportion of datasets satisfying $p_c \geq \alpha$ for 1000 evenly spaced values of $\alpha$ in $(0.01,0.99)$ when $\max(p_a,p_b) < \alpha$ and $p_d \geq \alpha$ under LSE-$F$ framework is shown in Figure \ref{fig: LSE-F1} (B). The significance level $\alpha$, which is the probability of the study rejecting the null hypothesis given that the null hypothesis is true, is often chosen to be smaller than 0.1 in practice and the proportion of the cases where $p_c > \alpha$ is larger than $40\%$, which indicates a high probability the total-effect test erroneously rejecting indirect-only mediation.

The proportion of erroneous total-effect test results for both directionally complementary and directionally competitive indirect-only mediation cases is depicted in Figure \ref{fig: LSE-F2}. The plot demonstrates that the total-effect test can lead to incorrect conclusions regarding the presence of indirect-only mediation in both cases. Interestingly, the erroneous judgments are more frequent when the signs of $\hat{a}\hat{b}$ and $\hat{d}$ are opposite, which is in line with expectations established in the analysis of Theorem \ref{thm: theorem1}.

% {\red 
% As mentioned before, indirect-only mediation consists of two sub-types, directionally complementary and directionally competitive, which correspond to the cases where $\hat{a}\hat{b}\hat{d} > 0$ and $\hat{a}\hat{b}\hat{d} \leq 0$. Similar to the above analysis, the proportion of erroneous total-effect test in both cases are shown in Figure \ref{fig: LSE-F2}. 
% It implies that the total-effect test can lead to wrong judgment of indirect-only mediation in both cases, and it is, not surprisingly, more frequent when $\hat{a}\hat{b}$ and $\hat{d}$ bear opposite signs.

\subsection{Exploratory analysis for other frameworks}
To investigate whether a similar result holds for other frameworks in establishing indirect-only mediation, we conducted a similar analysis using the LSE-Sobel framework and LAD-$Z$ framework with the same set of simulated datasets. Under the LSE-Sobel framework, we additionally calculated the the p-value $p_{ab}$ for the Sobel test of $a\times b$.
% we calculated the LSEs $(\hat{a},\hat{b},\hat{c},\hat{d})$ for each simulated dataset, and obtained the $p$-values $(p_c,p_d)$ for the $F$ test, as well as the p-value $p_{ab}$ for the Sobel test of $a\times b$.
If a similar result holds, 
% for the LSE-Sobel framework, 
we could expect to see $\{p_c \geq \alpha\} \not= \emptyset$ for any fixed $\alpha \in (0,1)$ when $p_{ab} < \alpha$ and $p_d \geq \alpha$, which is supported by Figures \ref{fig: LSE-Sobel1} and \ref{fig: LSE-Sobel2}. Moreover, graphical verification of results under the LAD-$Z$ framework are shown in Figures \ref{fig: LAD-Z1} and \ref{fig: LAD-Z2}, supporting similar conclusions.

% Similar to Figure \ref{fig: LSE-F1}, Figure \ref{fig: LSE-Sobel1} (A) shows that when $p_{ab} < \alpha = 0.1$ and $p_d \geq \alpha$, $\{p_c \geq \alpha\}$ is not empty. Additionally, Figure \ref{fig: LSE-Sobel1} (B) displays the proportion of datasets satisfying $p_c \geq \alpha$ for 1000 different values of $\alpha$ in $(0.01,0.99)$ when $p_{ab} < \alpha$ and $p_d \geq \alpha$ under LSE-Sobel framework. It's evident that a similar result holds for the LSE-Sobel framework as well. Figure \ref{fig: LSE-Sobel2} (A) and (B) show the probability of misleading total-effect test under directionally complementary and competitive scenarios, respectively, leading to a similar conclusion as in the LSE-$F$ framework. Graphical verification of results under the LAD-$Z$ framework are shown in Figures \ref{fig: LAD-Z1} and \ref{fig: LAD-Z2}, supporting similar conclusions.

\begin{figure}[t!]
    \centering
    \includegraphics[scale=0.55]{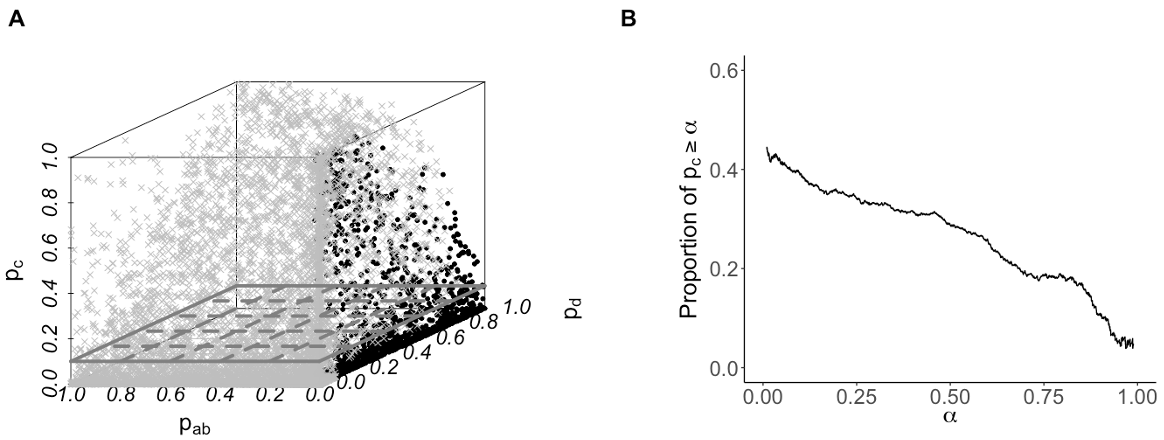}
    \caption{Numerical exploration under LSE-Sobel framework: (A) 
    black solid circles for $p_{ab} < 0.1$ and $p_d \geq 0.1$, grey crossings for $p_{ab} \geq 0.1$ or $p_d < 0.1$, 
    and dark gray dashed plane for $p_c = 0.1$;
    (B) Proportion of datasets satisfying $p_c \geq \alpha$ when $p_{ab} < \alpha$ and $p_d \geq \alpha$.
    % (A) Scatter plot of $p$-values with $\alpha = 0.1$ under LSE-Sobel framework: black solid circles for $p_{ab} < \alpha$ and $p_d \geq \alpha$, grey crossings for $p_{ab} \geq \alpha$ or $p_d < \alpha$, and dark gray dashed plane for $p_c = \alpha$. (B) Proportion of datasets satisfying $p_c \geq \alpha$ when $p_{ab} < \alpha$ and $p_d \geq \alpha$ under the LSE-Sobel framework.
    }
    \label{fig: LSE-Sobel1}
\end{figure}

\begin{figure}[t!]
    \centering
    \includegraphics[scale=0.25]{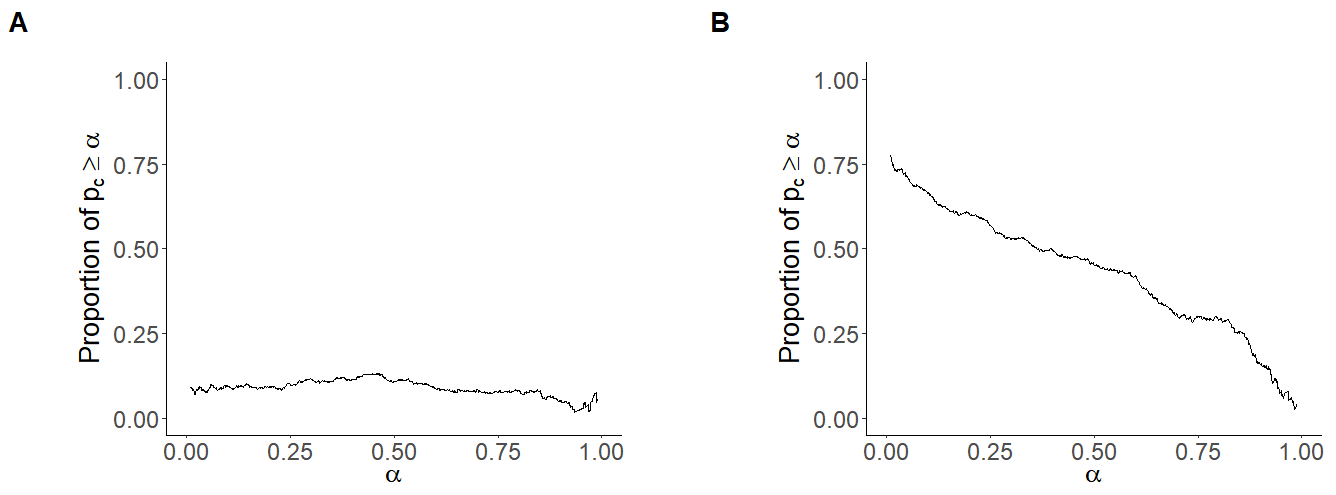}
    \caption{Proportion of datasets satisfying $p_c \geq \alpha$ when $p_{ab} < \alpha$, $p_d \geq \alpha$ and (A) $\hat{a}\hat{b}\hat{d} > 0$, (B) $\hat{a}\hat{b}\hat{d} < 0$ under the LSE-Sobel framework.}
    \label{fig: LSE-Sobel2}
\end{figure}

% Subfigure (A) shows that when $\max(p_a,p_b) < \alpha = 0.1$ and $p_d \geq \alpha$, the set $\{p_c \geq \alpha\}$, which consists of  solid circles above the dark gray dashed plane, is not empty. When $\max(p_a,p_b) < \alpha$ and $p_d \geq \alpha$, the proportion of datasets satisfying $p_c \geq \alpha$ for the 1000 different values of $\alpha$ is shown in Figure \ref{fig: LAD-Z1} (B) and it's easy to see that the cases where $p_c, p_d \geq \alpha$ while $\max(p_a,p_b) < \alpha$ is quite common. Figure \ref{fig: LAD-Z2} plots the proportion of insignificant total-effect test under two sub-types of indirect-only mediation and the properties are similar to that under LSE-$F$ and LSE-Sobel frameworks.

\begin{figure}[t!]
    \centering
    \includegraphics[scale=0.55]{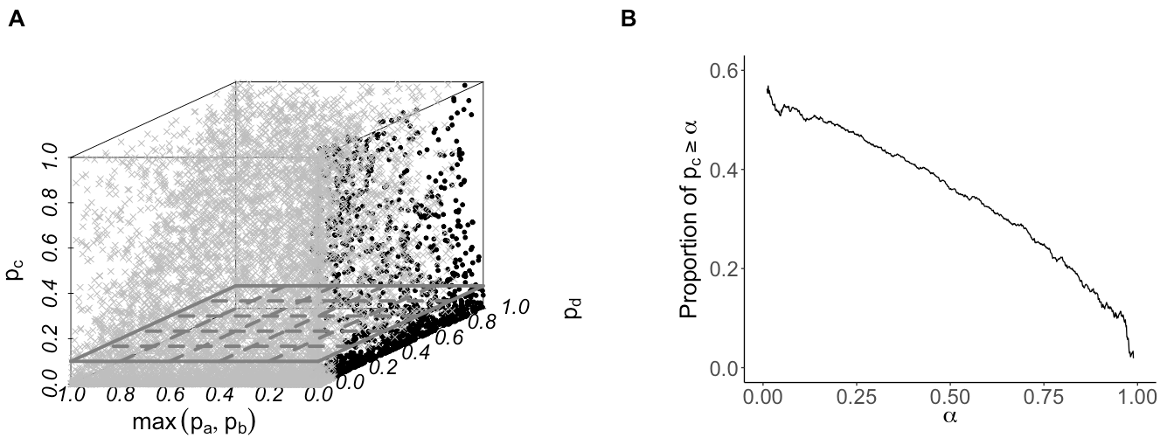}
    \caption{
    Numerical exploration under LAD-$Z$ framework: (A)  
    black solid circles for $\max(p_a,p_b) < 0.1$ and $p_d \geq 0.1$, grey crossings for $\max(p_a,p_b) \geq 0.1$ or $p_d < 0.1$, 
    and dark gray dashed plane for $p_c = 0.1$;
    (B) Proportion of datasets satisfying $p_c \geq \alpha$ when $\max(p_a,p_b) < \alpha$ and $p_d \geq \alpha$.
    % (A) Scatter plot of $p$-values with $\alpha = 0.1$ under LAD-$Z$ framework: black solid circles for $\max(p_a,p_b) < \alpha$ and $p_d \geq \alpha$, grey crossings for $\max(p_a,p_b) \geq \alpha$ or $p_d < \alpha$, 
    % and dark gray dashed plane for $p_c = \alpha$.
    % (B) Proportion of datasets satisfying $p_c \geq \alpha$ for different $\alpha$ when $\max(p_a,p_b) < \alpha$ and $p_d \geq \alpha$ under LAD-$Z$ framework.
    }
    \label{fig: LAD-Z1}
\end{figure}

\begin{figure}[t!]
    \centering
    \includegraphics[scale=0.25]{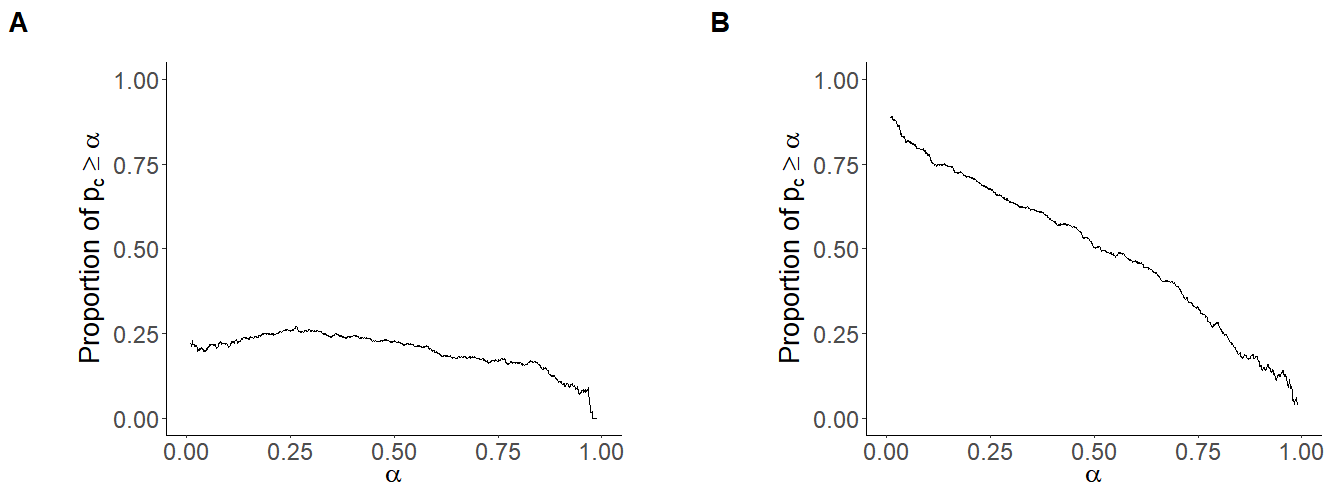}
    \caption{Proportion of datasets satisfying $p_c \geq \alpha$ when $\max(p_a,p_b) < \alpha$, $p_d \geq \alpha$ and (A) $\hat{a}\hat{b}\hat{d} > 0$, (B) $\hat{a}\hat{b}\hat{d} < 0$ under LAD-$Z$ framework.}
    \label{fig: LAD-Z2}
\end{figure}

\section{Real Data Illustrations} We illustrate the conclusions of the mathematical derivation and simulation presented above with two real-data examples. The example data came from the Health Information National Trends Survey (HINTS, \url{http://hints.cancer.gov/}), which is conducted regularly by the National Cancer Institute on representative samples of United States adults to track changes in health behavior and communication \citep{jiang2020digital,liu2023communication,finney2020data}. This study used the 2020 version of the postal-mail survey (HINTS 5 Cycle 4) with 3,865 participants.

Two models are presented below. Model 1 is a directionally competitive indirect-only (d-petitive IO) mediation depicting an effect of \textit{caregiving} (CG) on \textit{smoking} (SM) through \textit{psychological distress} (PD) (Figure \ref{fig: Real1}). Model 2 is a directionally complementary indirect-only (d-plementary IO) mediation describing the effect of \textit{employment} (EM) on \textit{physical activity} (PA) through \textit{psychological distress} (PD) (Figure \ref{fig: Real2}). In each model, the mediated effect passed the statistical threshold ($p<.05$) while the total effect failed ($p\geq .05$). 

In both examples, the total-effect ($c$) test would have concluded there was no ``effect to be mediated", which would be equal to ``no mediated effect" in the causal-steps doctrine, thereby requiring a full-stop of all further analysis, when the mediated effect ($ab$) would have passed the statistical test. Thus, each model is a real-data example of the total-effect test erroneously rejecting indirect-only mediation.

The two example models fit the definition of ``full mediation" aka ``complete mediation" under the quasi typology of  {full, partial, and no mediation} \cite{baron1986}.  The terms were meant to connote the strongest form of mediation \cite[p. 126]{Hayes2022IntroMediation}.  Hopefully, the total-effect test erroneously rejecting the perceived strongest mediation demonstrates the pitfalls of the total-effect test and the pitfalls of the causal-steps approach.

\begin{figure}[t!]
    \centering    \includegraphics[scale=0.55]{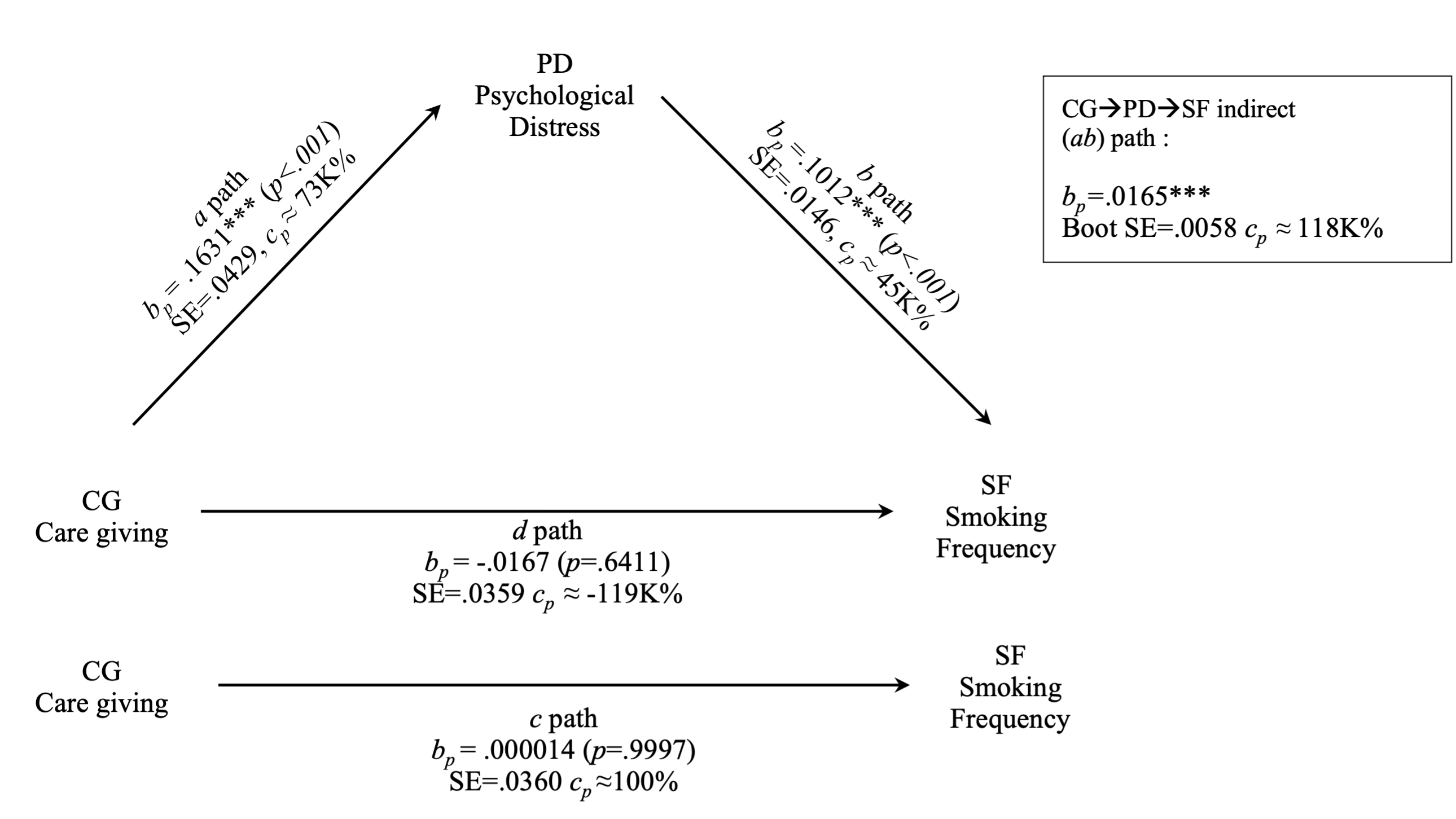}
    \caption{Directionally Competitive Indirect-only (D-petitive IO) Mediation ($n$ = 3,267)}
    %\protect\footnotemark[1]}
    \label{fig: Real1}
\end{figure}
% \footnotetext[1]{All variables are in percentage scales (0$\sim$1). Age, income, and education are controlled. *$p<.05$. **$p<.01$. ***$p<.001$.}

\begin{figure}[t!]
    \centering
    \includegraphics[scale=0.55]{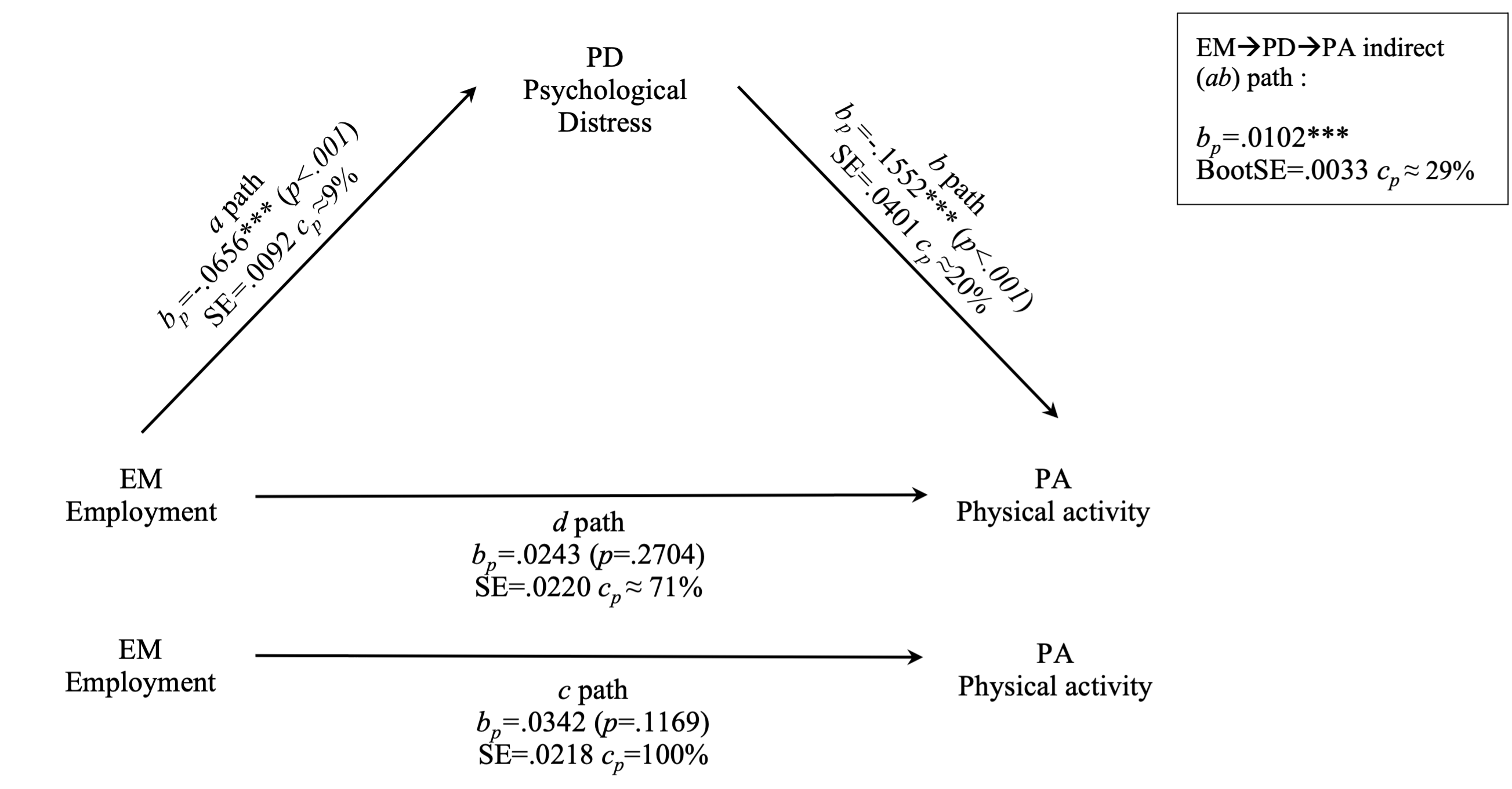}
    \caption{Directionally Complementary Indirect-only (D-plementary IO) Mediation ($n$ = 3,594)}
    %\protect\footnotemark[2]}
    \label{fig: Real2}
\end{figure}
% \footnotetext[2]{All variables are in percentage scales (0$\sim$1). Age, gender, and education are controlled. *$p<.05$. **$p<.01$. ***$p<.001$.}

When presenting the examples, we employ process-and-product analysis (PAPA) emerging in several disciplines \citep{jiang2021total,liu2023effect, Peng20203rdperson, Liu2023Electronic,zhao2014emerging,zhao1994media,zhao2010reconsidering}. PAPA approach sees mediation as a process and total effect (\textit{c}) as the product of the process. While the causal-steps approach is focused on one task, which is to ``establish mediation", PAPA is given three tasks, 1) testing effect hypotheses, 2) classifying effect types, and 3) analyzing effect sizes, all for the ultimate mission of better understanding the relationship between parts, process, and product.

To estimate effect sizes, PAPA employs the percentage coefficient ($b_p$), the regression coefficient with the dependent variable (DV) and independent variable (IV) both on a $0\sim 1$ percentage scale ($p_s$) \citep{jiang2021total,liu2022effects,zhao2016enough, Liu2023Electronic}. Scaled such, $b_p$ indicates the percentage change in DV associated with a 100$\%$ whole-scale increase in IV or, in other words, the change in DV measured by the percentage of a point associated with an increase in IV by one percentage point. Thus, $b_p$  is interpretable and comparable assuming scale equivalence  \citep{zhao2014emerging,jiang2021total}. The two features make $b_p$ a generic indicator of effect sizes that is easy to interpret and efficient to compare. Table \ref{table:1} provides scale details and univariate descriptions of variables, and Eq. 1 of Table \ref{table:2} is the formula for percentizing the scales.

\begin{table}[t!] %***
\caption{Descriptive statistics of variables in the HINTS Data: the sample size $N$, the original scale as data were collected and the 0-1 percentage scale after the variables have been linearly transformed to the interval $[0, 1]$.}
\label{table:1}\par
\resizebox{\linewidth}{!}{\begin{tabular}{|c c c c c c c c c c c c c c c c|} 
\hline
& & & \multirow{2}{*}{$N$} &  \multicolumn{4}{c}{Non-01 Natural Scale} & & \multicolumn{2}{c}{Conceptual Range} & & \multicolumn{4}{c|}{0-1 Percentage Scale} \\ 
\cline{5-8}\cline{10-11}\cline{13-16}
% \cmidrule(lr){5-8}  \cmidrule(lr){9-12}
& & & & Min & Max & Mean & SD & & Min & Max & & Min & Max & Mean & SD\\
 \hline
  \multirow{6}{4em}{Model 1} & $Y$ & Smoking Frequency & 3821 & 0 & 3 & 0.37 & 0.58 & & 0 & 3 & & 0 & 1 & 0.12 & 0.19\\
 & $M$ & Psychological Distress & 3810 & 0 & 4 & 0.67 & 0.97 & & 0 & 4 & & 0 & 1 & 0.17 & 0.24\\
 & $X$ & Caregiving & 3738 & 0 & 5 & 0.18 & 0.46 & & 0 & 5 & & 0 & 1 & 0.04 & 0.09\\
 & Control 1 & Age & 3738 & 18 & 104 & 57.01 & 16.99 & & 0 & 100 & & 0.18 & 1.04 & 0.57 & 0.17\\
 & Control 2 & Income & 3448 & 1 & 9 & 5.59 & 2.26 & & 1 & 9 & & 0 & 1 & 0.57	& 0.28\\
 & Control 3 & Education & 3722 & 1 & 7 & 4.94 & 1.62 & & 1 & 7 & & 0 & 1 & 0.66 & 0.27\\
  \hline
 \multirow{6}{4em}{Model 2} & $Y$ & Physical Activity & 3739 & 0 & 4620 & 160.09 & 273.43 & & 0 & 500 & & 0 & 9.24 & 0.03 & 0.06\\
 & $M$ & Psychological Distress & 3810 & 0 & 4 & 0.67 & 0.97 & & 0 & 4 & & 0 & 1 & 0.17 & 0.24\\
 & $X$ & Employment & 3778 & 0 & 1 & 0.50 & 0.50 & & 0 & 1 & & 0 & 1 & 0.50 & 0.50\\
 & Control 1 & Age & 3738 & 18 & 104 & 57.01 & 16.99 & & 0 & 100 & & 0.18 & 1.04 & 0.57 & 0.17\\
 & Control 2 & Gender & 3765 & 1 & 2 & 1.59 & 0.49 & & 1 & 2 & & 0 & 1 & 0.59 & 0.49\\
 & Control 3 & Education & 3722 & 1 & 7 & 4.94 & 1.62 & & 1 & 7 & & 0 & 1 & 0.66 & 0.27\\
   \hline
  \end{tabular}}
\end{table}

\begin{table}[t!] %***
\caption{Percentage coefficient ($b_p$) and percent contribution ($c_p$) to total effect ($c$).}
\label{table:2}\par
\resizebox{\linewidth}{!}{\begin{tabular}{|c c c c c c c c c|} 
 \hline
\multicolumn{3}{|c}{Indicator} & \multicolumn{3}{c}{$\text{Equation}^1$} & \multicolumn{2}{c}{Range} & Eq.\\
 \hline
\multicolumn{3}{|c}{\makecell[l]{Percentage score ($p_s$)}}
 & \multicolumn{3}{c}{$p_s = \frac{o_s - c_{min}}{c_{max}-c_{min}}$} & \multicolumn{2}{c}{\makecell[c]{Observable $-\infty < p_s < \infty$\\Conceptual $0 < p_s < 1$}} & 1\\
\hline 
\multicolumn{3}{|c}{\makecell[l]{Percent contribution of total effect\\ to total effect ($c$)}} & \multicolumn{3}{c}{$c_p(c) = \frac{b_p(c)}{|b_p(c)|}$} & \multicolumn{2}{c}{$c_p(c) = 1$ or $-1$} & 2\\
\hline 
\multicolumn{3}{|c}{\makecell[l]{Percent contribution of indirect \\ effect ($ab$) to total effect ($c$)}} & \multicolumn{3}{c}{$c_p(ab) = \frac{b_p(ab)}{|b_p(c)|}$} & \multicolumn{2}{c}{\makecell[c]{$-\infty < c_p(ab) < \infty$\\ $|c_p(ab)| \leq |c_p(c)|$}} & 3\\
\hline 
\multicolumn{3}{|c}{\makecell[l]{Percent contribution of direct $\&$\\ remainder effect ($d$) to total effect ($c$)}} & \multicolumn{3}{c}{$c_p(d) = \frac{b_p(d)}{|b_p(c)|}$} & \multicolumn{2}{c}{\makecell[c]{$-\infty < c_p(d) < \infty$\\ $|c_p(d)| \leq |c_p(c)|$}} & 4\\
\hline 
\multicolumn{3}{|c}{\makecell[l]{Percent contribution of $\text{1}^{\text{st}}$-leg effect\\ ($a$) to total effect ($c$)}} & \multicolumn{3}{c}{$c_p(a) = \frac{|b_p(a)|}{|b_p(a)|+|b_p(b)|} \times c_p(ab)$} & \multicolumn{2}{c}{\makecell[c]{$-\infty < c_p(a) < \infty$\\ $|c_p(a)| \leq |c_p(ab)|$}} & 5\\
\hline 
\multicolumn{3}{|c}{\makecell[l]{Percent contribution of $\text{2}^{\text{nd}}$-leg effect\\ ($b$) to total effect ($c$)}} & \multicolumn{3}{c}{$c_p(b) = \frac{|b_p(b)|}{|b_p(a)|+|b_p(b)|} \times c_p(ab)$} & \multicolumn{2}{c}{\makecell[c]{$-\infty < c_p(b) < \infty$\\ $|c_p(b)| \leq |c_p(ab)|$}} & 6\\
\hline 
\multicolumn{3}{|l}{For Eq. 1} & \multicolumn{3}{l}{For Eqs. 2$\sim$6} & \multicolumn{3}{l|}{For Eqs. 2$\sim$6 (continued)}\\
\multicolumn{3}{|l}{$c_{max}$: conceptual maximum} & \multicolumn{3}{l}{$(a)$: first leg of the indirect path.} & \multicolumn{3}{l|}{$b_p$: percentage coefficient, regression}\\
\multicolumn{3}{|l}{$c_{min}$: conceptual minimum} & \multicolumn{3}{l}{$(b)$: second leg of the indirect path.} & \multicolumn{3}{l|}{\ \ \ \ \ coefficient when DV and IV is each}\\
\multicolumn{3}{|l}{$o_s$: original score} & \multicolumn{3}{l}{$(c)$: total effect.} & \multicolumn{3}{l|}{\ \ \ \ \ on a percentage scale ($p_s$).}\\
\multicolumn{3}{|l}{$p_s$: percentage score} & \multicolumn{3}{l}{$(d)$: direct $\&$ remainder path.} & \multicolumn{3}{l|}{$c_p$: percent contribution to total effect, $c$.}\\
& & & \multicolumn{3}{l}{$(ab)$: indirect path.} & & &\\
\hline 
\multicolumn{9}{|l|}{\makecell[l]{Note 1: Concepts and equations adopted and adapted from \cite{jiang2021total}, \cite{liu2023communication}, \cite{zhao2014emerging}\\ and \cite{zhao2016enough}.}}\\
\hline 
\end{tabular}}
\end{table}

To help dissect the product, discern the parts and divine the process, PAPA calculates \textit{percent contribution} ($c_p$), the contribution of each elemental part, i.e., the $a$, $b$, $ab$, $d$, or $c$ path, to the $IV \rightarrow DV$ total effect, $c$, as detailed in Table \ref{table:2}. To reduce overuse and misuse of $p$ values, we strive to practice what we consider the best practice, 1) refraining from the term “statistical significance” and “statistical non-significance”; 2) referring to $p<.05$ as “statistical acknowledgment”, \citep{benjamini2021asa,nature2019s,siegfried2015p,wilkinson1999statistical} and 3) referring to $p\geq .05$ as “statistical inconclusiveness”. Such practices indicate that $p<.05$ is merely a pretest yardstick or partition threshold under the principles of functionalism, passing which would allow for classifying  effect types and analyzing effect sizes \citep{liu2023covid,zhao2022interrater,liu2023effect}.  It's hoped that such practices, including the application of effect-size indicators such as $b_p$ and $c_p$, benefit from and contribute to the ``effect size movement''  \citep{Kelley2012effectsize,preacher2011effect, wilkinson1999statistical, robinson2002effect,jiang2021total,Schmidt1996significance}. 

\subsection{Model 1 for 
%Directionally Competitive Indirect-Only (
D-petitive IO Mediation}

Competitive mediation, aka suppression or inconsistent mediation \citep{mackinnon2000equivalence,mackinnon2007mediation}, defined as a model with statistically acknowledged $ab$ and $d$ paths at the opposite directions, is widely known as the type of mediation that can be erroneously rejected by the total-effect test \citep{xiao2018social,busse2016abc,gopalakrishnan2019client}. The mathematical derivation above has provided proof that the total-effect test can also erroneously reject indirect-only (IO) mediation, which includes two subtypes, directionally competitive (d-petitive) and directionally complementary (d-plementary). The following is a real-data example for the first subtype, D-petitive IO mediation, defined as a model with a statistically acknowledged $ab$ path and a statistically inconclusive $d$ path in the opposite direction. The example shows that this model of mediation is erroneously rejected by the total-effect test. 

\noindent\textbf{Key Variables.}\\
\noindent\textbf{Dependent variable:}
\textit{Smoking frequency} (SF) was measured by four items that asked responders how often they smoke cigarettes and e-cigarettes. The composite variable ranged from 0 to 1 where 0 represents not smoking and 1 represents smoking every day.\\
\noindent\textbf{Mediating variable:}
\textit{Psychological distress} (PD) was the sum of four items (Cronbach’s $\alpha= .871$) measuring the frequency by which the respondents experienced four symptoms of psychological distress in the past two weeks, feeling little interest in doing things, being emotionally down, hopeless, and anxious. Again, it was transferred into a $0 \sim 1$ where 0 means never feeling any symptoms and 1 means feeling four symptoms every day. \\
\noindent\textbf{Independent variable:}
\textit{Caregiving (CG)} was the sum of five items, measuring whether the respondents took five types of caregiving currently. The caregiving types include caring for children, partners, parents, relatives, and friends. The compound variable CG ranges $0 \sim 1$ where 0 indicates no caregiving responsibilities and 1 indicates the respondent needs to take all types of caregiving.

\noindent\textbf{Control variables:}
\textit{Age}, \textit{income}, and \textit{education} were reported in Table \ref{table:1} and included as control variables for the mediation analysis. To simplify the presentation, the control variables were omitted from Table \ref{table:3} and Figures \ref{fig: Real1} $\&$ \ref{fig: Real2} that report the outcomes.

\noindent\textbf{Mediation analysis.}\\
Table \ref{table:3} and Figure \ref{fig: Real1} summarize Model 1 findings. They show a positive and statistically acknowledged indirect effect, and a negative but statistically inconclusive direct-and-remainder (di-remainder) effect, producing a directionally competitive indirect-only (d-petitive IO) mediation. Figure \ref{fig: Real1} reports the contribution of each path. The $ab$ path ($b_p=.0165\%$, $c_p\approx 117,857\%$) and $d$ path ($b_p=-.0167$, $c_p\approx -119,286\%$) contributed about equal percentages in opposite directions, leading to the nearly-zero total effect ($c$ path, $b_p=.000014$). Because the competition between $ab$ and $d$ paths was about equal (118K$\%$ v.s. -119K$\%$), they offset each other to produce a near-zero total effect ($b_p=.000014$). Consequently, the contribution of each part appeared huge percentage-wise. In such cases of small total effect due to even competition, the comparative sizes may be more important than the sizes themselves. 

The indirect path ($ab$) passed the statistical threshold ($p<.001$) while the direct-and-remainder ($d$) path failed ($p=.6411$), making it a directionally competitive indirect-only (d-petitive IO) mediation by our standards \citep{zhao2011does,zhao2010reconsidering,hayes2009beyond,rucker2011mediation}. Nevertheless, the total effect ($c$) failed ($p=.9997$). If passing the total-effect test remains a necessary condition for establishing mediation as some experts continued to prescribe, this model would have been disqualified as mediation \citep{wen2004testing,wen2014analyses,rose2004mediator}.

\subsection{Model 2 for
%Directionally Complementary Indirect-Only (
D-plementary IO Mediation}
In the above example, the competition between the indirect ($ab$) and the direct-and-remainder ($d$) paths was clearly a main contributor to the near-zero total effect ($c$) and the statistical inconclusiveness. The following is a real-data example that a second subtype, a directionally complementary indirect-only (d-plementary IO) mediation, is erroneously rejected by the total-effect test, thereby showing that the total-effect test can erroneously reject mediation without competition.

\noindent\textbf{Key Variables.}\\
\noindent\textbf{Dependent variable:}
\textit{Physical activity} (PA) was measured by two items that asked responders how many minutes per day and how many days per week they usually did physical activity \citep{xie2020electronic,kontos2014predictors}. The two items were multiplied to compute the weekly physical activity the respondents conducted. The composite variable was then linearly transformed to a 0-1 percentage scale where 1 represents the highest weekly physical activity and 0 represents not conducting weekly physical activity at all.

\noindent\textbf{Mediating variable:}
\textit{Psychological distress} (PD) in Model 2 was the same mediating variable in Model 1.

\noindent\textbf{Independent variable:}
\textit{Employment} (EM) measured whether the respondents were employed or not, e.g., unemployed, retired, or being students, recoded 1 for employed and 0 for not employed.

\noindent\textbf{Control variable:}
\textit{Age}, \textit{gender}, and \textit{education} were controlled in mediation analysis but omitted from Table \ref{table:3} and Figures \ref{fig: Real1} $\&$ \ref{fig: Real2} that report the outcomes.

\noindent\textbf{Mediation analysis.}\\
Model 2 of Table \ref{table:3} show that the indirect ($ab$) path was positive and statistically acknowledged, while the dire-mainder ($d$) path was also positive but failed to pass the statistical threshold ($p=.2704$), making it a directionally complementary indirect-only (d-plementary IO) mediation. The total effect ($c$) also failed the statistical test ($p=.1169$), demonstrating that total-effect test can also erroneously reject this subtype of mediation. This could be among the first documented real-data examples that the total-effect test erroneously rejects mediation without competition, aka suppression. 

A question arises: Given that the indirect path was statistically acknowledged, the direct-and-remainder path complemented, and the total effect was the sum of the two paths, what caused the statistical inconclusiveness of the total effect?

The mathematical derivations provided above would point fingers at the estimated variance (SE) of the $d$ path. That is, for this subtype, the large variance of the $d$ path relative to the other parameters should be considered the largest factor contributing to the larger-than-threshold $p$-value of the total effect ($c$). The process-and-product analysis (PAPA) of the real-data example provides a non-mathematical illustration of the point.    

As shown in Figure \ref{fig: Real2}, even though the $d$ path was statistically inconclusive ($p=.2704$), the estimated effect size of the $d$ path ($b_p=.0243$) more than doubled that of the statistically acknowledged $ab$ path ($b_p=.0102$, $p<.001$). The contribution of the d path accounted for 71$\%$ of the total effect ($c_p=71\%$). The effect size of the $d$ path is not small relative to the other main parameters. Rather, it was the relatively large variance of the $d$ path (SE = .022) that was a main contributor to the variance of the $c$ path (SE = .0218), which was, in turn, a main factor contributing to the statistical inconclusiveness of the total effect ($p=.1169$).

\subsection{Section Summary and Discussion}
The previous sections proved a mathematical theorem and provided Monte Carlo simulations showing that the total-effect test can erroneously reject indirect-only (IO) mediation. This section added two real-data examples documenting that the test did erroneously reject this type of mediation. Two models were shown, recording erroneous rejection for each of the two major subtypes of IO mediation. Model 1 is a case where the total-effect test erroneously rejected directionally competitive indirect-only (d-petitive IO) mediation. Model 2 is a case where the total-effect test erroneously rejected directionally complementary (d-plementary IO) mediation.

As the d-plementary IO model involves no competition, Model 2 shows that the erroneous rejection can occur without competition, statistical or directional. The finding has implications for data analysts who struggle to interpret d-plementary IO models with statistically inconclusive total effects. It also illustrates an implication of Theorem \ref{thm: theorem1}: The estimation variance, i.e., standard error, of direct-and-remainder ($d$) path tends to be large relative to other parameters of this subtype; the estimation variance of the direct-and-remainder path may be the more significant factor than effect sizes and other parameters contributing to the statistical inconclusiveness of the total effect ($c$).  

An implication is that practicing researchers may need to focus more on effect sizes than on \textit{p}-values or confidence intervals. For example, in a d-plementary IO model, the statistically inconclusive $p$-value for the $c$ path is attributed more to the large variance in the $d$ path.  While the $d$ portion of the $c$ path can not be acknowledged, the direction of the $c$ path still can, due mainly to the direction and strength of the $ab$ path. The direction and strength of the effect are often theoretically and practically as important as, if not more important than, the variance of the effect.

\begin{table}[t!] %***
\caption{Mediation analysis results of the HINTS Data. Columns 2-5: LSEs of the parameters.}
\label{table:3}\par
\resizebox{\linewidth}{!}{\begin{tabular}{| c c c c c c c c c c c|} 
 \hline
\multirow{2}{*}{Model}
&  \multicolumn{4}{c}{Estimates} & \multirow{2}{*}{$I(\hat{a}\hat{b}\hat{d}>0)$} & \multicolumn{4}{c}{$p$-values} & \multirow{2}{*}{Mediation Type} \\ 
% \cmidrule(lr){4-7}\cmidrule(lr){9-12}
\cline{2-5}\cline{7-10}
& $a$ & $b$ & $d$ & $c$ & & $p_a$ & $p_b$ & $p_d$ & $p_c$ & \\
 \hline
\multirow{2}{*}{Model 1} & \multirow{2}{*}{.1631} & \multirow{2}{*}{.1012} & \multirow{2}{*}{$-.0167$} & \multirow{2}{*}{.000014} & \multirow{2}{*}{No} & \multirow{2}{*}{$<.001$} & \multirow{2}{*}{$<.001$} & \multirow{2}{*}{$.6411$} & \multirow{2}{*}{$.9997$} & \multirow{2}{*}{\makecell[c]{Directionally Competitive\\  Indirect-only Mediation}}\\
& & & & & & & & & & \\
% {\makecell[c]{Directionally \\ Competitive \\  Indirect-only \\ Mediation}}\\
% & $M$ & Psychological distress & & & & & & & & & &\\
% & $X$ & Employment & & & & & & & & & &\\
% & Control 1 & Age & & & & & & & & & &\\
% & Control 2 & Gender & & & & & & & & & &\\
% & Control 3 & Education & & & & & & & & & &\\
\hline 
\multirow{2}{*}{Model 2} & \multirow{2}{*}{$-.0656$} & \multirow{2}{*}{$-.1552$} & \multirow{2}{*}{$.0243$} & \multirow{2}{*}{.0342} & \multirow{2}{*}{Yes} & \multirow{2}{*}{$<.001$} & \multirow{2}{*}{$<.001$} & \multirow{2}{*}{$.2704$} & \multirow{2}{*}{$.1169$} & 
\multirow{2}{*}{\makecell[c]{Directionally Complementary\\  Indirect-only Mediation}}\\
& & & & & & & & & & \\
% \multirow{1}{*}{\makecell[c]{Directionally \\ Complementary\\ Indirect-only\\ Mediation}}\\
% & $M$ & Psychological distress & & & & & & & & & &\\
% & $X$ & Caregiving & & & & & & & & & &\\
% & Control 1 & Age & & & & & & & & & &\\
% & Control 2 & Income & & & & & & & & & &\\
% & Control 3 & Education & & & & & & & & & &\\
\hline 
\multicolumn{11}{|l|}{\makecell[l]{Note: See Table S1 in the Supplementary Material for variables in Models 1 and 2. As all variables are on 0-1 percentage scales,\\ all regression coefficients, namely $\hat{a}$, $\hat{b}$, $\hat{d}$, and $\hat{c}$, become percentage coefficients ($b_p$).}}\\
\hline 
  \end{tabular}}
\end{table}

\section{Conclusions and Discussions}
\subsection{Main Findings}
This study provided a mathematical theorem, a {Monte Carlo} simulation, and two real-data examples to demonstrate that the total-effect test can and did erroneously reject indirect-only mediation. There are three and only three types of mediation, competitive, complementary, and indirect-only \citep{busse2016abc,jiang2021total,zhao2011does,zhao2010reconsidering}. While prior studies have shown that the total-effect test 
% can erroneously reject competitive mediation and 
is superfluous for establishing complementary mediation, this study shows that the test can erroneously reject indirect-only and competitive mediation. 
Thus, this study completes the argument that the total-effect test should be recanted, but not just relaxed or suspended, for establishing any type of mediation, to the extent that the traditional ordinary least square (LSE-$F$ or LSE-Sobel) procedures were applied to calculate $p$-values or confidence intervals for statistical tests. 
A similar conclusion can be reached under the
% {\red LSE-Sobel} and 
LAD-$Z$ framework, which is verified with simulation studies.

Table \ref{table:role&contributors} displays the repercussions of imposing the total-effect test for establishing mediation. Altogether three types of mediation  and the three types of tests make up the nine cells. The table shows two possible types of outcomes, i.e., repercussions, to be erroneous or superfluous. The total-effect test is erroneous in seven of the nine situations and is superfluous in the other two situations. 

The table also lists the studies that have contributed to the knowledge and influenced this study the most directly. Of all the nine cells, ``this study" is the only entry in three (C2, B3, C3), indicating it is the only contributor.  ``This study" also appears in two other cells (B1 and C1). If we count the nine cells as nine pieces of knowledge, this study emerges as the sole or main contributor to five of the nine pieces regarding the repercussions of the test. 

% \begin{table}[t!] %***
% \caption{Role of total-effect test for different mediation types and the main contributors.}
% \label{table:role&contributors}\par
% \resizebox{\linewidth}{!}{\begin{tabular}{|c c| c|c|c |} 
%  \hline
% \multicolumn{2}{|c|}{\multirow{2}{*}{\diagbox{Framework}{Mediation Type}}}& \multirow{2}{*}{Complementary}  & \multirow{2}{*}{Competitive}  & \multirow{2}{*}{Indirect-Only} \\
% & & & & \\
% \hline
% \multicolumn{2}{|c|}{\multirow{3}{*}{LSE-$F$}}
%  & Superfluous & Contradiction	 & Contradiction	\\
%  \cline{3-5}
%  & & \cite{zhao2011does} & \citet{kenny1998data,jiang2021total} & \cite{jiang2021total}\\
%   & & \cite{jiang2021total}
%   % \cite{zhao2010reconsidering}
%   & This paper & This paper\\
%   \hline
%   \multicolumn{2}{|c|}{\multirow{3}{*}{LSE-Sobel}}
%  & Superfluous & Contradiction	 & Contradiction	\\
%  \cline{3-5}
%  & & \citet{zhao2010reconsidering,zhao2011does} & \citet{mackinnon2000equivalence,zhao2010reconsidering}  & \cite{zhao2011does}\\
%   & & \cite{jiang2021total}
%   & This paper & This paper \\
%   \hline
%     \multicolumn{2}{|c|}{\multirow{3}{*}{LAD-$Z$}}
%  & Contradiction	 & Contradiction	 & Contradiction	\\
%  \cline{3-5}
%  & & \cite{jiang2021total} & This paper & This paper \\
%   & & 
%   % \cite{zhao2010reconsidering}
%   % & \cite{rucker2011mediation} & \cite{zhao2011does} \\
%   & & \\
%  \hline
%   \end{tabular}}
% \end{table}

\begin{table}[t!] %***
\caption{Repercussions of Total-Effect Test and Contributors to Knowledge}
\label{table:role&contributors}\par
\resizebox{\linewidth}{!}{\begin{tabular}{|c |c| c|c|c|} 
 \hline
\multirow{3}{*}{\makecell[c]{Down: \\  Type of \\ Statistical Test}} & \multirow{3}{*}{\makecell[c]{Right: \\  Type of \\ Mediation Models}} & \multirow{3}{*}{\makecell[c]{A\\Complementary / \\ Partial  Mediation}}  & \multirow{3}{*}{\makecell[c]{B\\Competitive / Inconsistent / \\ Suppresive Mediation}}  & \multirow{3}{*}{\makecell[c]{C\\Indirect-Only / \\Full  Mediation}} \\
& & & & \\
& & & & \\
\hline
\multirow{4}{*}{1. LSE-$F$} & Test outcomes$^1$ & { \textit{\textbf{Superfluous}}} & { \textit{\textbf{Erroneous}}} & { \textit{\textbf{Erroneous}}}\\
\cdashline{2-5}[3pt/3pt]
& \multirow{3}{*}{Contributing studies$^2$} & \multirow{3}{*}{\makecell[c]{\cite{zhao2010reconsidering}\\ \cite{jiang2021total}}} & \multirow{3}{*}{\makecell[c]{\cite{kenny1998data}\\ \cite{jiang2021total}\\ This study$^3$}} & \multirow{3}{*}{\makecell[c]{\cite{jiang2021total}\\ This study$^3$}}\\
& & & & \\
& & & & \\
\hline
\multirow{4}{*}{2. LSE-Sobel} & Test outcomes$^1$ & { \textit{\textbf{Superfluous}}} & { \textit{\textbf{Erroneous}}} & { \textit{\textbf{Erroneous}}}\\
\cdashline{2-5}[3pt/3pt]
& \multirow{3}{*}{Contributing studies$^2$} & \multirow{3}{*}{\makecell[c]{\cite{zhao2010reconsidering}\\ \cite{jiang2021total}}} & \multirow{3}{*}{\makecell[c]{\cite{mackinnon2000equivalence}\\ \cite{hayes2009beyond}\\ \cite{zhao2010reconsidering}}} & \multirow{3}{*}{This study$^3$}\\
& & & & \\
& & & & \\
\hline
\multirow{2}{*}{3. LAD-$Z$} & Test outcomes$^1$ & { \textit{\textbf{Erroneous}}} & { \textit{\textbf{Erroneous}}} & { \textit{\textbf{Erroneous}}}\\
\cdashline{2-5}[3pt/3pt]
& Contributing studies$^2$ & \cite{jiang2021total} & This study$^3$ & This study$^3$\\
\hline
\multicolumn{5}{|l|}{\makecell[tl]{1. The three rows show two possible outcomes of total-effect tests, i.e., erroneous or superfluous. Cell A1, \\ for example, shows that, with complementary mediation using LSE-$F$ test, the total effect test is superfluous, \\ because it would always pass the statistical threshold to ``establish'' this type of mediation. For another example, \\ Cell A2 shows that the total-effect test would erroneously reject competitive mediation using LSE-$F$ test, \\with ``mediation'' defined by the near-consensus that the \textit{ab} path passes the $p < \alpha$ statistical test \\ ($\alpha$=0.05 for this study).\\  2. This table is not meant to provide a comprehensive survey of past studies or a ranking of contributors. \\ We instead selected a maximum of three studies per cell that have had the most direct impact on this study.\\
3. For Cells C2, B3, and C3, this study is the main contributor to the conclusion that the total-effect test can \\ be erroneous for establishing mediation.}}\\
\hline
  \end{tabular}}
\end{table}
Now, this study has provided proof that the total-effect test can produce erroneous outcomes, not only for competitive mediation but also for indirect-only mediation, both directionally competitive and complementary. Every cell of Table \ref{table:role&contributors} has been filled. The total-effect test harms or fails to help, whether LSE-$F$, LSE-Sobel, or LAD-$Z$ test is used. The burden of proof is now on the causal-step procedure to show that the total-effect test is harmless and helpful for establishing any type of mediation. 

At the meantime, it might be appropriate to consider recanting the total-effect test for establishing mediation of all types, rather than imposing the test and then suspending or relaxing it for one special type. This means completely removing the test from the regular procedure instead of allowing exceptions in the cases of anticipated suppression or inconsistency, aka competitive mediation \citep{baron1986,kenny1998data,kenny2008reflections,kenny2021mediation,rose2004mediator,wen2004testing,wen2014analyses}.

Admittedly, proofs assuming bootstrap tests are not yet available. But the available evidence, especially if adding this study, is overwhelmingly against the test. Given the striking imbalance between pro and con, it's time to set aside the total-effect test for establishing mediation unless and until new evidence emerges to show a benefit.

\subsection{Implications} 

Why and how did the total-effect test dominate so many disciplines for so long \citep{baron1986}?  Over-dichotomization of the \textit{effect} concept and oversimplification of the \textit{mediation} concept may be among the root causes. In light of the revelation, it may be time to revisit and possibly revamp the objectives, concepts, theories, and techniques of mediation analysis. It may be time also to consider more broadly \textit{causal dissection models}, which include moderation and curvilinearity in addition to mediation \citep{zhao2017MainEffect}. Accordingly, Section 6 above are tasked to showcase two applications of \textit {process-and-product analysis} (PAPA), whose missions are 1) testing the presence of mediation, 2) identifying types of mediation, and 3) analyzing sizes of the mediation and non-mediation effects \citep{jiang2021total,liu2022effects,Liu2023Electronic}. Establishing mediation is a part, and only a part, of the first mission of PAPA.

\subsection{Future Research} 

While this study completes the argument about recanting total-effect test for establishing mediation using OLS tests, i.e., LSE-$F$, LSE-Sobel, and LAD-$Z$, future research may falsify or qualify the argument by testing the main theses assuming non-OLS tests such as bootstraps \citep{zhao2011does}. The more challenging task is to overcome the spiral of inertia, resist over-dichotomization and oversimplification of fundamental concepts, and foster an understanding of mediation that is more analytical and comprehensive. It  will take time and luck, but above all persistence. (c.f. \cite{zhao2022interrater}; \cite{zhao2018WeAgreed})

%%%%%%%%%%%%%%%%%%%%%%%%%%%%%%%%%%%%%%%%%%%%%%%%%%%%%%%%%%%%%%%%%%%%%%%%%%%%%%%%%%%%%%%%%%%%%%%%%%%%%%%%%%%%%%%%%%%%%%%%%%%%
\section*{Supplementary Material}

\textit {Supplementary Material} is attached at the end of this document to provide the detailed proof for Lemma \ref{lem:InvarianceUnderOrthogonalTransformation}, \ref{lem:LSE-F|Sobel}, \ref{lem:competitive}, Theorem \ref{thm:SobelComplementary}, \ref{thm:SobelIO}, \ref{thm: competitive} and \ref{thm:SobelCompetitive} and results of 
simulation for competitive mediation under LSE-$F$, LSE-Sobel, and LAD-$Z$ frameworks. 
% The document also provides more detailed information about the real-data examples, the historical context of the debate over the total-effect test, and the implications of this study's findings.
\par

%%%%%%%%%%%%%%%%%%%%%%%%%%%%%%%%%%%%%%%%%%%%%%%%%%%%%%%%%%%%%%%%%%%%%%%%%%%%%%%%%%%%%%%%%%%%%%%%%%%%%%%%%%%%%%%%%%%%%%%%%%%%
\section*{Acknowledgments}

This research was supported by the Beijing Natural Science Foundation [Z190021]; National Natural Science Foundation of China [Grant 11931001];
grants of University of Macau, including CRG2021-00002-ICI, ICI-RTO-0010-2021, CPG2021-00028-FSS and SRG2018-00143-FSS, ZXS PI; and Macau Higher Education Fund, HSS-UMAC-2020-02, ZXS PI. Xinshu Zhao and Ke Deng are co-corresponding authors.

% \par

%%%%%%%%%%%%%%%%%%%%%%%%%%%%%%%%%%%%%%%%%%%%%%%%%%%%%%%%%%%%%%%%%%%%%%%%%%%%%%%%%%%%%%%%%%%%%%%%%%%%%%%%%%%%%%%%%%%%%%%%%%%%

\par

%%%%%%%%%%%%%%%%%%%%%%%%%%%%%%%%%%%%%%%%%%%%%
\bibhang=1.7pc
\bibsep=2pt
\fontsize{9}{14pt plus.8pt minus .6pt}\selectfont
\renewcommand\bibname{\large \bf References}
%\begin{thebibliography}{11}
\expandafter\ifx\csname
natexlab\endcsname\relax\def\natexlab#1{#1}\fi
\expandafter\ifx\csname url\endcsname\relax
  \def\url#1{\texttt{#1}}\fi
\expandafter\ifx\csname urlprefix\endcsname\relax\def\urlprefix{URL}\fi

% \newpage
%% use bibfile 
% \bibliographystyle{chicago}      % Chicago style, author-year citations
% \bibliography{bib}   % name your BibTeX data base

% %%  Another method
% \begin{thebibliography}{}

% \bibitem[Curtis(1943)]{1943}
% Curtis, M. (1943).
% {\em Documents on International Affairs, 1938}, Volume~II.
%  Oxford University Press, London.

% \bibitem[Eubank(2004)]{Eubank}
% Eubank, K. (2004).
%  {\em The origins of World War II}. 
%  3rd Edition.
% Harlan Davidson, Wheeling, Ill.

% \bibitem[Gellately(1988)]{1988}
% Gellately, R. (1988).
%  The gestapo and german society: Political denunciation in the gestapo
%   case files.
% {\em The Journal of Modern History}~{\bf 60}, 654--694.

% \bibitem[Noakes and Pridham(2001)]{Noakes}
% Noakes, J. and G.~Pridham (2001).
%  {\em Nazism, 1919-1945. Vol. 3: Foreign Policy, War and Racial
%   Extermination}.
% University of Exeter Press,  Exeter.

% \end{thebibliography}

%%%%%%%%%%%%%%%%%%%%%%%%%%%%%%%%%%%%%%%%%%%%%%%%%%%%%%%%%%%%%%%%%%%%%%%%%%%%%%%%%%%%%%%%%%%%%%%%%%%%%%%%%%%%%%%%%%%%%%%%%%%%
\vskip .65cm
\noindent
Tingxuan Han\\
Center for Statistical Science \& Department of Industry Engineering, Tsinghua University, Haidian, Beijing, China.
\vskip 2pt
\noindent
E-mail: htx21@mails.tsinghua.edu.cn
\vskip 2pt

\noindent
Luxi Zhang
\vskip 2pt \noindent
Department of Communication, University of Macau, Macau, China.
\vskip 2pt \noindent
E-mail: yc27303@um.edu.mo

\noindent
Xinshu Zhao
\vskip 2pt \noindent
Department of Communication, University of Macau, Macau, China.
\vskip 2pt \noindent
E-mail: xszhao@um.edu.mo

\noindent
Ke Deng\\
Center for Statistical Science \& Department of Industry Engineering, Tsinghua University, Haidian, Beijing, China.
\vskip 2pt
\noindent
E-mail: kdeng@tsinghua.edu.cn

\newpage
\appendix

\begin{center}
    \textbf{\centering \Large Supplementary Material}
\end{center}

This supplementary material contains two sections. Section 1 (S1) presents the 
detailed proof for Lemma \ref{lem:InvarianceUnderOrthogonalTransformation}, \ref{lem:LSE-F|Sobel}, \ref{lem:competitive}, Theorem \ref{thm:SobelComplementary},
\ref{thm:SobelIO}, \ref{thm: competitive} and \ref{thm:SobelCompetitive}.
Section 2 (S2) shows the results of simulation for competitive mediation under LSE-$F$, LSE-Sobel and LAD-$Z$ frameworks. 
% Section 3 (S3) provides more details of the two real-data examples. Section 4 (S4) provides a more comprehensive review of the discussion and debate over several decades on whether the total-effect test should be required for establishing mediation, thereby placing this study in the theoretical and historical context. 
\par

\setcounter{section}{0}
\setcounter{table}{0}
\setcounter{equation}{0}
\setcounter{figure}{0}
\def\theequation{S\arabic{section}.\arabic{equation}}
\def\thefigure{S\arabic{section}.\arabic{figure}}
\def\thesection{S\arabic{section}}
\def\thelemma{S\arabic{lemma}}
\def\thetable{S\arabic{table}}
\fontsize{12}{14pt plus.8pt minus .6pt}\selectfont

\section{Technical Proofs}
\subsection{Proof of Lemma \ref{lem:InvarianceUnderOrthogonalTransformation}}
The invariance of $F$-test has been proved in \cite{jiang2021total} and we focus on the invariance of Sobel test here.
To simplify the problem,
consider the following general multivariate linear regression problem:
\[\begin{split}
    Y = \beta_0 X_0 + \beta_1 X_1 + \cdots + \beta_p X_p + \varepsilon
\end{split}\]
with $n$ observed data points $\{(X_{i0},X_{i1},\ldots,X_{ip},Y_i)\}$. Define the vector of coefficients
$\bbeta = (\beta_0,\beta_1,\ldots,\beta_p)'$, design matrix $\bX = (\bX_0,\bX_1,\ldots,\bX_p)$, where $\bX_j = (X_{1j},\ldots,X_{nj})'$, and response vector $\bY = (Y_1,\ldots,Y_n)'$. Denote the 2-norm of a vector as $||\cdot||_2^2$. Then the least square estimator of $\bbeta$ takes the form of 
\[\begin{split}
    \hat{\bbeta} = (\bX'\bX)^{-1}\bX'\bY,
\end{split}\]
and the covariance is 
\[\begin{split}
    {\bbcov}(\hat{\bbeta}) = s^2 (\bX'\bX)^{-1},
\end{split}\]
where $s^2 = ||\bY-\bY_{\bX}||_2^2/(n-p-1)$, $\bY_{\bX}$ represents projection of $\bY$ onto the space spanned by $\bX$. As Sobel test is based on $\hat{\bbeta}$ and ${\bbcov}(\hat{\bbeta})$, it suffices to show that for orthogonal matrix $\Gamma$ and constant $\gamma$, the LSE and estimated covariance of coefficient of the regression problem with transformed data matrix $(\Tilde{\bX}_0,\ldots,\Tilde{\bX}_p,\Tilde{\bY}) = \gamma\Gamma(\bX_0,\ldots,\bX_p,\bY)$ remain unchanged.

First, the invariance of LSE is easy to see as
\[\begin{split}
\Tilde{\bbeta} = (\Tilde{\bX}'\Tilde{\bX})^{-1}\Tilde{\bX}'\Tilde{\bY} = (\bX'\Gamma'\Gamma\bX)^{-1}\bX'\Gamma'\Gamma\bY = (\bX'\bX)^{-1}\bX'\bY = \hat{\bbeta}.
\end{split}\]
Moreover, since 
\[\begin{split}
    ||\Tilde{\bY}-\Tilde{\bY}_{\Tilde{\bX}}||_2^2 = ||\gamma\Gamma(\bY-\bY_{\bX})||_2^2 = \gamma^2 ||\bY-\bY_{\bX}||_2^2,
\end{split}\]
and 
\[\begin{split}
    (\Tilde{\bX}'\Tilde{\bX})^{-1} = (\gamma^2 \bX'\Gamma'\Gamma\bX)^{-1} = (\gamma^2 \bX'\bX)^{-1},
\end{split}\]
we have 
\[\begin{split}
{\bbcov}(\Tilde{\bbeta}) = ||\Tilde{\bY}-\Tilde{\bY}_{\Tilde{\bX}}||_2^2/(n-2)\cdot
(\Tilde{\bX}'\Tilde{\bX})^{-1} = s^2(\bX'\bX)^{-1} = {\bbcov}(\hat{\bbeta}).
\end{split}\]
Above all, the LSE and estimated covariance are invariant under orthogonal transformation. Hence, the invariance of LSE-Sobel test holds as well.

\subsection{Proof of Lemma \ref{lem:LSE-F|Sobel}}

By definition, the test statistics in Sobel test is 
\[\begin{split}
    S = \frac{\hat{a}\hat{b}}{\left(\hat{a}^2 \bbV(\hat{b}) + \hat{b}^2 \bbV(\hat{a})\right)^{1/2}} = \frac{T_aT_b}{\sqrt{T_a^2+T_b^2}},
\end{split}\]
where $T_a^2 = \hat{a}^2/\bbV(\hat{a})$ and $T_b^2 = \hat{b}^2/\bbV(\hat{b})$. Based on the transformed data matrix $\Tilde{\mathcal{D}}$, we have 
$$\hat{a} = \frac{m_2}{x_2},\ \ \hat{b} = \frac{y_3}{m_3},\ \ \bbV(\hat{a}) = \frac{m_3^2}{(n-2)x_2^2}, \ \ \bbV(\hat{b}) = \frac{y_4^2}{(n-3)m_3^2}.$$
Hence, $T_a = \sqrt{n-2}\cdot m_2/m_3$, $T_b = \sqrt{n-3}\cdot y_3/y_4$, and $$|S| = \frac{1}{(1/T_a^2+1/T_b^2)^{1/2}} = \frac{1}{\{1/[(n-2)r^2] + 1/[(n-3)p^2]\}^{1/2}}.$$
As sample size $n\rightarrow \infty$, $S$ follows standard normal distribution asymptotically. Hence, the rejection region of $a\times b$ is 
\[\begin{split}
    \mathcal{R}_{a\times b}(\alpha) = \left\{|S| > z_{\alpha/2}\right\} = \left\{\frac{1}{(n-2)r^2} + \frac{1}{(n-3)p^2} < \frac{1}{z_{\alpha/2}^2}\right\}.
\end{split}\]

\subsection{Proof of Theorem \ref{thm:SobelComplementary}}
\setcounter{equation}{0}
As the geometric plots of rejection regions have been shown in \cite{jiang2021total}, we only give the mathematical analysis here.

Let $\mathcal{R}_{\beta}(\alpha \mid r)$ be the intersection of $\mathcal{R}_{\beta}(\alpha)$ and the $p$-$q$ plane $\mathcal{P}_r$ for all $\beta \in \{a,b,d,c, a\times b\}$.
It is straightforward to see that $\mathcal{R}_{\beta}(\alpha) = \bigcup_r \mathcal{R}_{\beta}(\alpha \mid r)$, and 
\[\begin{split}
    \Bar{\cR}_c(\alpha) \cap \mathcal{R}_{a\times b}(\alpha) \cap {\cR}_d(\alpha) = \bigcup_{r>0}\left\{
    \Bar{\cR}_c(\alpha\mid r) \cap \mathcal{R}_{a\times b}(\alpha\mid r) \cap {\cR}_d(\alpha\mid r)\right\}.
\end{split}\]
As implied by Lemma \ref{lem:LSE-F|Sobel},   $\mathcal{R}_{a\times b}(\alpha|r) = \{p>p_0(r)\}$, where 
\begin{equation}\label{eq:p0_r2}
    \frac{1}{(n-2)r^2} + \frac{1}{(n-3)p_0^2(r)} =\frac{1}{z_{\alpha/2}^2},\ (n-2)r^2 > z_{\alpha/2}^2.
\end{equation}
% We show that $\Bar{\mathcal{R}}_{c}(\alpha|r) \cap \mathcal{R}_{a\times b}(\alpha|r) \cap \mathcal{R}_{d}(\alpha|r) \rightarrow \emptyset$ for any $r\geq 0$.
% When $r \leq z_{\alpha/2}/\sqrt{n-2}$, it is easy to see that $\mathcal{R}_{a\times b}(\alpha|r) = \emptyset$ and the argument holds. Therefore, we only need to consider the case where $r > z_{\alpha/2}/\sqrt{n-2}$.
By \cite{jiang2021total}, $\hat{a}\hat{b}\hat{d}>0$ implies $\hat{a}\hat{b}\hat{c}>0$ and $\cR_d(\alpha\mid r) = \{q>rp + p_{n,\alpha}(1+r^2)^{1/2}\}$.
Then for $r^2 > z_{\alpha/2}^2/(n-2)$, the intersection of $\mathcal{R}_{a\times b}(\alpha\mid r)$, ${\cR}_d(\alpha\mid r)$ and $\Bar{\cR}_c(\alpha\mid r)$ is
\[\begin{split}
    \left\{p>p_0(r),  rp + p_{n,\alpha}(1+r^2)^{1/2} < q \leq r_{n,\alpha}{(p^2+1)}^{1/2} \right\}.
\end{split}\]
Hence, it suffices to show that as $n\rightarrow \infty$, \begin{equation}\label{eq:target2}
P\left(\cD\in \bigcup_{r>r_0(n)}\left\{p>p_0(r), rp + p_{n,\alpha}(1+r^2)^{1/2}<q \leq r_{n,\alpha}{(p^2+1)}^{1/2}\right\}\right) \rightarrow 0,
    \end{equation}
where $r_0(n) = z_{\alpha/2}/\sqrt{n-2}$. 

Since the probability density function of $F$-distribution with degrees of freedom $(1,n)$ satisfies 
\[\begin{split}
    f_{F_{1,n}}(x) &= \frac{\Gamma(n/2+1/2)}{\sqrt{\pi}\Gamma(n/2)}\cdot \frac{1}{\sqrt{nx}}\left(1+\frac{x}{n}\right)^{-\frac{n+1}{2}}\rightarrow \frac{1}{\sqrt{2\pi x}}e^{-x/2}
\end{split}\]
as $n\rightarrow \infty$, we have $\lambda_{n-2}(\alpha) \rightarrow \chi^2_1(\alpha)$, where $\chi^2_1(\alpha)$ is the $\alpha$th-quantile of $\chi^2_1$ distribution. Hence, $\sqrt{\lambda_{n-2}(\alpha)} \rightarrow \sqrt{\chi^2_1(\alpha)} \equiv z_{\alpha/2}$, i.e., $\sqrt{n-2}\cdot r_{n,\alpha} \equiv \sqrt{\lambda_{n-2}(\alpha)} \rightarrow z_{\alpha/2} = \sqrt{n-2}\cdot r_0(n)$. Therefore, 
% \[\begin{split}
%     &\ \ \  \lim_{n\rightarrow\infty} P\left(\cD \in \{r_0(n)<r < r_{n,\alpha}\}\right)=0.
% \end{split}\]
$    \{r: r_0(n)<r < r_{n,\alpha}\} \rightarrow \emptyset.$
According to \cite{jiang2021total},
\[\begin{split}
    \bigcup_{r\geq r_{n,\alpha}}\left\{p>0: rp + p_{n,\alpha}{(1+r^2)}^{1/2}<r_{n,\alpha}{(p^2+1)}^{1/2}\right\}= \emptyset.
\end{split}\]
Hence, 
\[\begin{split}
    &\ \ \ P\left(\cD\in \bigcup_{r>r_0(n)}\left\{p>p_0(r), rp + p_{n,\alpha}{(1+r^2)}^{1/2}<r_{n,\alpha}{(p^2+1)}^{1/2}\right\}\right)\\&= P\left(\cD\in \bigcup_{r_0(n)<r<r_{n,\alpha}}\left\{p>p_0(r), rp + p_{n,\alpha}{(1+r^2)}^{1/2}<r_{n,\alpha}{(p^2+1)}^{1/2}\right\}\right)\\
    &\rightarrow 0,
\end{split}\]
and the argument \eqref{eq:target2} holds. 

\subsection{Proof of Lemma \ref{lem:competitive}}
% It suffices to show that $q < rp$.
By Lemma \ref{lem:LSE-F|Sobel}, 
\[\begin{split}
    \hat{a}\hat{b}\hat{d} = \frac{m_2y_3(m_3y_2-m_2y_3)}{x_2^2m_3^2},\ \ \hat{a}\hat{b}\hat{c} =\frac{m_2y_2y_3}{x_2^2m_3}.
\end{split}\]
Since $x_2>0$ and $m_3 > 0$, conditions $\hat{a}\hat{b}\hat{d} < 0$ and $\hat{a}\hat{b}\hat{c} \geq 0$ imply that
\[\begin{split}
    m_2m_3y_2y_3 < m_2^2y_3^2\text{ and }
    m_2y_2y_3 \geq 0.
\end{split}\]
Hence, $m_2m_3y_2y_3 > 0$ and $|m_3y_2| < |m_2y_3|$, which is equivalent to $q < rp$ as $y_4 > 0$.

\subsection{Proof of Theorem \ref{thm:SobelIO}}
We show that there exists $N>0$ such that for any $n>N$, we can find some $r_n>0$ such that $\Bar{\mathcal{R}}_{c}(\alpha|r_n) \cap \mathcal{R}_{a\times b}(\alpha|r_n) \cap \Bar{\cR}_{d}(\alpha|r_n) \not= \emptyset$. 
When $\hat{a}\hat{b}\hat{c} \geq 0$, $\mathcal{R}_{a\times b}(\alpha\mid r)\cap\Bar{\cR}_c(\alpha\mid r)\cap\Bar{\cR}_d(\alpha\mid r)$ takes the form  
$$\left\{\max\big\{rp- p_{n,\alpha}(r^2+1)^{1/2},0\big\} \leq q \leq \min\{r_{n,\alpha}(p^2+1)^{1/2},rp+ p_{n,\alpha}(r^2+1)^{1/2}\},p>p_0(r)\right\},$$
and when $\hat{a}\hat{b}\hat{c}<0$, the above intersection is $$\left\{0 \leq q \leq \min\big\{r_{n,\alpha}(p^2+1)^{1/2},-rp+ p_{n,\alpha}(r^2+1)^{1/2}\big\},p>p_0(r)\right\}.$$
% As $\mathcal{R}_{a\times b}(\alpha|r) = \emptyset$  for $r \leq z_{\alpha/2}/\sqrt{n-2}$, we only consider the case where $r > z_{\alpha/2}/\sqrt{n-2}$.
It is easy to see that $\Bar{\mathcal{R}}_{c}(\alpha|r) \cap \mathcal{R}_{a\times b}(\alpha|r) \cap \Bar{\cR}_{d}(\alpha|r) \not= \emptyset$ holds if $p_0(r) < p_{n,\alpha}(r^2+1)^{1/2}/r.$
As the definition of $p_0(r)$ implies $r > z_{\alpha/2}/\sqrt{n-2}$,
it suffices to show that there exists $N>0$ such that for any $n>N$, we can find some 
% $r > z_{\alpha/2}/\sqrt{n-2}$
$r_n > z_{\alpha/2}/\sqrt{n-2}$
such that 
% \begin{equation}\label{eq:targetIO}
% \left\{p > p_0(r_n): g_1(p) \equiv r_{n,\alpha}\sqrt{p^2+1}-r_n p+p_{n,\alpha}\sqrt{1+r_n^2} >0\right\}\not=\emptyset.
% \end{equation}
\begin{equation}\label{eq:targetIO2}
p_0(r_n) < p_{n,\alpha}{(r_n^2+1)}^{1/2}/r_n.
\end{equation}

Let $r_n^2 = 2z_{\alpha/2}^2/(n-2)$, then Eq.~\eqref{eq:p0_r2} implies $p_0^2(r_n) = 2z_{\alpha/2}^2/(n-3)$. Since $\sqrt{n-3}\cdot p_{n,\alpha} \rightarrow z_{\alpha/2}$ and $r_n^2 \rightarrow 0$, there exists $N > 0$ s.t. for $n>N$, we have $\sqrt{n-3}\cdot p_{n,\alpha} \geq z_{\alpha/2}/2$ and $r_n < 1/3$. Therefore, when $n > N$, we have $$\sqrt{n-3}\cdot p_{n,\alpha}{(r_n^2+1)}^{1/2}/r_n \geq \sqrt{10}\cdot z_{\alpha/2}/2>\sqrt{2}z_{\alpha/2} = \sqrt{n-3}\cdot p_0(r_n).$$

Above all, for $n>N$ and $r_n^2 = 2z_{\alpha/2}^2/(n-2)$, we have $\Bar{\mathcal{R}}_{c}(\alpha|r_n) \cap \mathcal{R}_{a\times b}(\alpha|r_n) \cap \Bar{\cR}_{d}(\alpha|r_n) \not= \emptyset$, and thus, $\Bar{\mathcal{R}}_{c}(\alpha) \cap \mathcal{R}_{a\times b}(\alpha) \cap \Bar{\cR}_{d}(\alpha) \not= \emptyset$. 

\subsection{Proof of Theorem \ref{thm: competitive}}

% \setcounter{equation}{0}
% \par
% Since $\mathcal{R}_a(\alpha\mid r) = \mathcal{P}_r \cap \{r>r_{n,\alpha}\}=\emptyset$ when $r \leq r_{n,\alpha}$, it suffices to show that $\mathcal{R}_a(\alpha\mid r) \cap \mathcal{R}_b(\alpha\mid r) \cap \mathcal{R}_d(\alpha\mid r) \cap \Bar{\cR}_c(\alpha\mid r) \not= \emptyset$ for some $r>r_{n,\alpha}$.

$\mathcal{R}_a(\alpha) \cap \mathcal{R}_b(\alpha) \cap \mathcal{R}_d(\alpha) \cap \Bar{\cR}_c(\alpha) \not= \emptyset$ can be equivalently expressed as 
$\mathcal{R}_a(\alpha\mid r) \cap \mathcal{R}_b(\alpha\mid r) \cap \mathcal{R}_d(\alpha\mid r) \cap \Bar{\cR}_c(\alpha\mid r) \not= \emptyset$ for some $r>r_{n,\alpha}$ since
$\mathcal{R}_a(\alpha\mid r) = \mathcal{P}_r \cap \{r>r_{n,\alpha}\}=\emptyset$ when $r \leq r_{n,\alpha}$. By definition, 
$\mathcal{R}_b(\alpha\mid r) = \{p>p_{n,\alpha}\}$, and $\Bar{\cR}_c(\alpha\mid r)=\left\{p \geq 0, 0 \leq q \leq r_{n,\alpha}(p^2+1)^{1/2}\right\}$. 
% corresponds to the space between the $q$-axis and the 
% higher branch of the hyperbola with asymptotes $q \pm r_{n,\alpha}p$ and vertices $(0,\pm r_{n,\alpha})$. 
When $\hat{a}\hat{b}\hat{c} < 0$, 
$$\mathcal{R}_d(\alpha\mid r)=\left\{p \geq 0, q > -rp+ p_{n,\alpha}(r^2+1)^{1/2}\right\},$$ and
% is the region above straight line $q = -rp+ p_{n,\alpha}(r^2+1)^{1/2}$. 
for $\hat{a}\hat{b}\hat{c} \geq 0$, Lemma \ref{lem:competitive} implies that $$\mathcal{R}_d(\alpha\mid r)=\left\{p \geq 0,0 \leq q < rp - p_{n,\alpha}(r^2+1)^{1/2}\right\}.$$
% is below the straight line $q = rp - p_{n,\alpha}(r^2+1)^{1/2}$.
\begin{figure}[t!]
    \centering
    \includegraphics[scale=0.07]{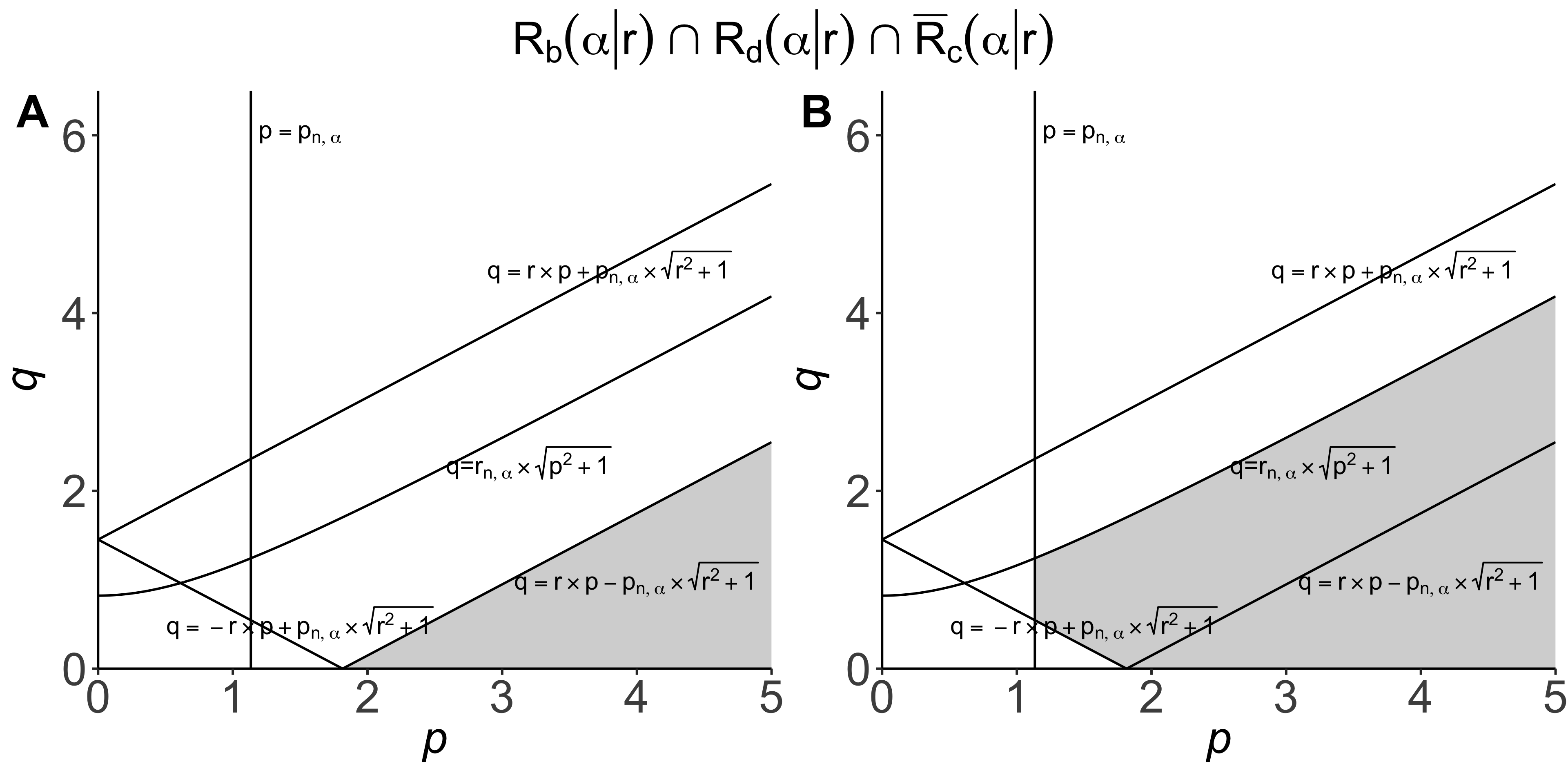}
    \caption{Geometry of $\mathcal{R}_a(\alpha\mid r) \cap \mathcal{R}_b(\alpha\mid r) \cap \mathcal{R}_d(\alpha\mid r) \cap \mathcal{R}_c^c(\alpha\mid r)$ in  $p\operatorname{-}q$ plane with $r > r_{n,\alpha}$ for competitive mediation when (A) $\hat{a}\hat{b}\hat{c} \geq 0$ and (B) $\hat{a}\hat{b}\hat{c} < 0$.}
    \label{fig: competitive2_2}
\end{figure}
The geometry of $\mathcal{R}_a(\alpha\mid r) \cap \mathcal{R}_b(\alpha\mid r) \cap \mathcal{R}_d(\alpha\mid r) \cap \Bar{\cR}_c(\alpha\mid r)$ is demonstrated in Figure \ref{fig: competitive2_2}.
% It is easy to check that the intersection is not empty when
% $p \in (p_{n,\alpha},p_{n,\alpha}(r^2+1)^{1/2}/r)$.
When $\hat{a}\hat{b}\hat{c} \geq 0$, it suffices to show that $r_{n,\alpha}(p^2+1)^{1/2} > \max\{0,rp - p_{n,\alpha}(r^2+1)^{1/2}\}$ for some $p > p_{n,\alpha}$, which always holds for $p \in (p_{n,\alpha},p_{n,\alpha}(r^2+1)^{1/2}/r)$. When $\hat{a}\hat{b}\hat{c} < 0$, it suffices to show that $r_{n,\alpha}(p^2+1)^{1/2} > 0$ for some $p>p_{n,\alpha}$, which always holds. 
We thereby conclude the proof of Theorem \ref{thm: competitive}.

{ As shown in Figure \ref{fig: competitive2_2}, the intersection of $\mathcal{R}_{a\times b}(\alpha)$, $ {\mathcal{R}}_{d}(\alpha)$ and $\Bar{\mathcal{R}}_{c}(\alpha)$ under $\hat{a}\times \hat{b} \times \hat{c} \geq  0$ is a subset of that under $\hat{a}\times \hat{b} \times \hat{c} < 0$, which implies the total-effect test is more likely to be erroneous for establishing competitive mediation when $\hat{a}\times \hat{b} \times \hat{c} < 0$.}

\subsection{Proof of Theorem \ref{thm:SobelCompetitive}}
We show that $\Bar{\mathcal{R}}_{c}(\alpha|r) \cap \mathcal{R}_{a\times b}(\alpha|r) \cap {\cR}_{d}(\alpha|r) \not= \emptyset$ for some $r$ when $\hat{a}\hat{b} \hat{d} < 0$. We only consider the case where $r > z_{\alpha/2}/\sqrt{n-2}$ since $\mathcal{R}_{a\times b}(\alpha|r) = \emptyset$ when this condition doesn't hold. The expression of $\mathcal{R}_d(\alpha|r)$ is the same as that under the LSE-$F$ framework.
% When $\hat{a}\hat{b}\hat{c} \geq 0$, Lemma \ref{lem:competitive} implies that $\mathcal{R}_d(\alpha|r) = \{p\geq 0,q < rp - p_{n,\alpha}(r^2+1)^{1/2}\}$, and when $\hat{a}\hat{b}\hat{c} < 0$, $\mathcal{R}_d(\alpha|r) = \{p\geq 0,q > -rp + p_{n,\alpha}(r^2+1)^{1/2}\}$. 
As $\Bar{\cR}_c(\alpha|r) = \{p\geq 0,q \leq r_{n,\alpha}(p^2+1)^{1/2}\}$ and $\cR_{a\times b}(\alpha|r) = \{p > p_0(r)\}$, the observation that $\Bar{\mathcal{R}}_{c}(\alpha|r) \cap \mathcal{R}_{a\times b}(\alpha|r) \cap {\cR}_{d}(\alpha|r) \not= \emptyset$ is trivial. 

\section{Simulations for Competitive Mediation}
\setcounter{equation}{0}

This section presents the results of Monte Carlo simulation  
% detailed proof for Lemma \ref{lem:InvarianceUnderOrthogonalTransformation}, \ref{lem:LSE-F|Sobel}, Theorem \ref{thm:SobelComplementary}, \ref{thm:SobelIO}, \ref{thm: competitive} and \ref{thm:SobelCompetitive}.
for competitive mediation under LSE-$F$, LSE-Sobel and LAD-$Z$ frameworks. 
% will be included as well.
\par
To validate 
Theorem \ref{thm: competitive}, we generate the simulated data from model \eqref{eq: model1} and \eqref{eq: model2} as follows:
\[\begin{split}
    n \sim \text{Unif}(\{10,\ldots,100\}),\quad (i_M,i_Y,a,b,d) \sim \text{Unif}[-1,1]^5,
\end{split}\]
\[\begin{split}
    X \sim N(0,1),\quad \sigma_M^2\text{ and } \sigma_Y^2\sim \text{Inv-Gamma}(1,1).
\end{split}\]
A total of $10,000$ independent datasets of different sample sizes were simulated. For each simulated dataset, the LSEs $(\hat{a},\hat{b},\hat{c},\hat{d})$ and their $p$-values $(p_a,p_b,p_c,p_d)$ under the LSE-$F$ framework are calculated. If Theorem \ref{thm: competitive} holds, then when for any fixed $\alpha \in (0,1)$, we have $\{p_c \geq \alpha\} \cap \{\hat{a}\hat{b}\hat{d} < 0\} \not= \emptyset$ if $\max(p_a,p_b,p_d) < \alpha$.

Figure \ref{fig: competitive_LSEF} (A) checks the $p$-value condition when $\alpha=0.1$ by demonstrating each simulated dataset with one point in a 2-dimensional space with $\max(p_a,p_b,p_d)$ and $p_c$ be the $X$ and $Y$-axis, respectively. The solid circles stand for datasets satisfying $\max(p_a,p_b,p_d) < \alpha$ and $\hat{a}\hat{b}\hat{d} < 0$, gray crossings represent data sets such that $\max(p_a,p_b,p_d) \geq \alpha$ or $\hat{a}\hat{b}\hat{d} \geq 0$, and the dark gray dashed line represents $p_c = \alpha$. Solid circles above the line $p_c = \alpha$ form an empirical version of set $\{p_c \geq \alpha\}$. We can see that when $\max(p_a,p_b,p_d) < \alpha = 0.1$, $\{p_c \geq \alpha\}\cap \{\hat{a}\hat{b}\hat{d} < 0\}$ is not empty. To check the theoretical result for different values of $\alpha$, the proportion of datasets satisfying $p_c \geq \alpha$ for 1000 evenly spaced values of $\alpha$ in $(0.01,0.99)$ when $\max(p_a,p_b,p_d) < \alpha$ and $\hat{a}\hat{b}\hat{d} < 0$
under LSE-$F$ framework is shown in Figure \ref{fig: competitive_LSEF} (B). It implies that the total-effect test will reject competitive mediation erroneously with quite large probability.

\begin{figure}[t!]
    \centering
    \includegraphics[scale=0.27]{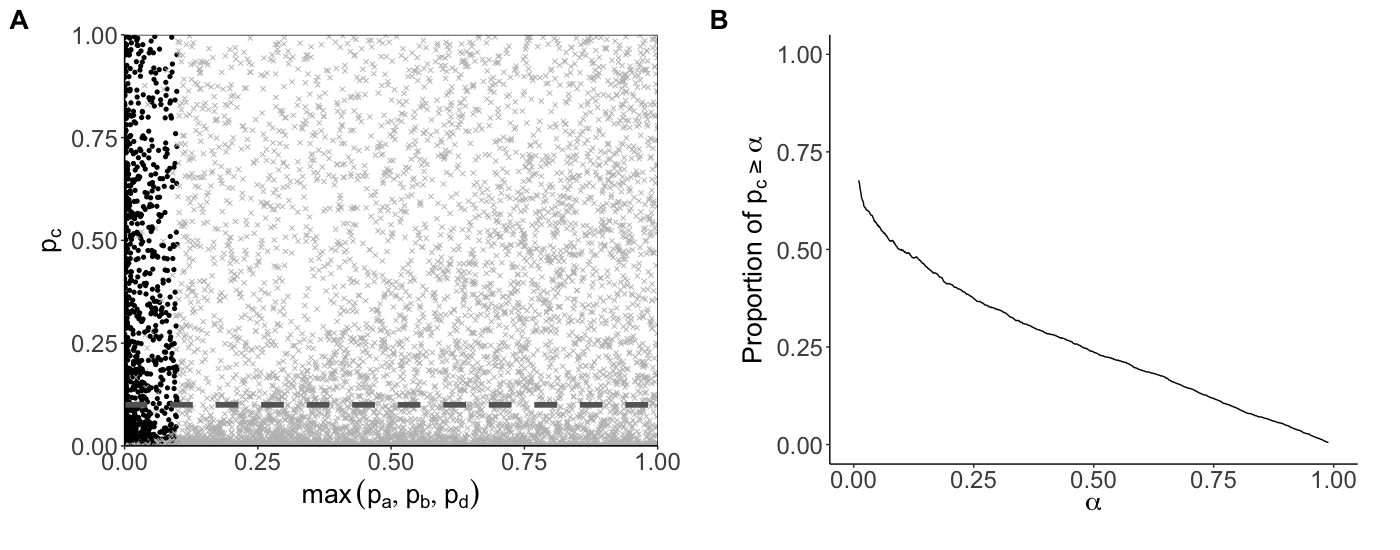}
    \caption{(A) Scatter plot of $p$-values with $\alpha = 0.1$ under LSE-$F$ framework: black solid circles represent datasets with $\max(p_a,p_b,p_d) < \alpha$ and $\hat{a}\hat{b}\hat{d} < 0$, grey crossings represent datasets with $\max(p_a,p_b,p_d) \geq \alpha$ or $\hat{a}\hat{b}\hat{d} \geq 0$,
    and the dark gray dashed line represents $p_c = \alpha$.
    (B) Proportion of datasets satisfying $p_c \geq \alpha$ for different $\alpha$ when $\max(p_a,p_b,p_d) < \alpha$ and $\hat{a}\hat{b}\hat{d} < 0$ under LSE-$F$ framework.}
    \label{fig: competitive_LSEF}
\end{figure}

Similar analysis under LSE-Sobel framework and LAD-$Z$ framework could be conducted to test whether a similar result holds for other frameworks for establishing competitive mediation. For LSE-Sobel framework, we calculated the LSEs $(\hat{a},\hat{b},\hat{c},\hat{d})$ for each simulated dataset using the same group of simulated datasets. $p$-values $(p_c,p_d)$ of $F$-test as well as the $p$-value $p_{ab}$ of Sobel test for $a \times b$ are calculated. If a similar result holds for LSE-Sobel framework, we could expect to see for any fixed $\alpha \in (0,1)$, when $\max(p_{ab},p_d) < \alpha$, we have $\{p_c \geq \alpha\} \cap \{\hat{a}\hat{b}\hat{d}<0\} \not= \emptyset$, which is supported by the results in Figure \ref{fig: competitive_LSESobel}.

\begin{figure}[t!]
    \centering
    \includegraphics[scale=0.27]{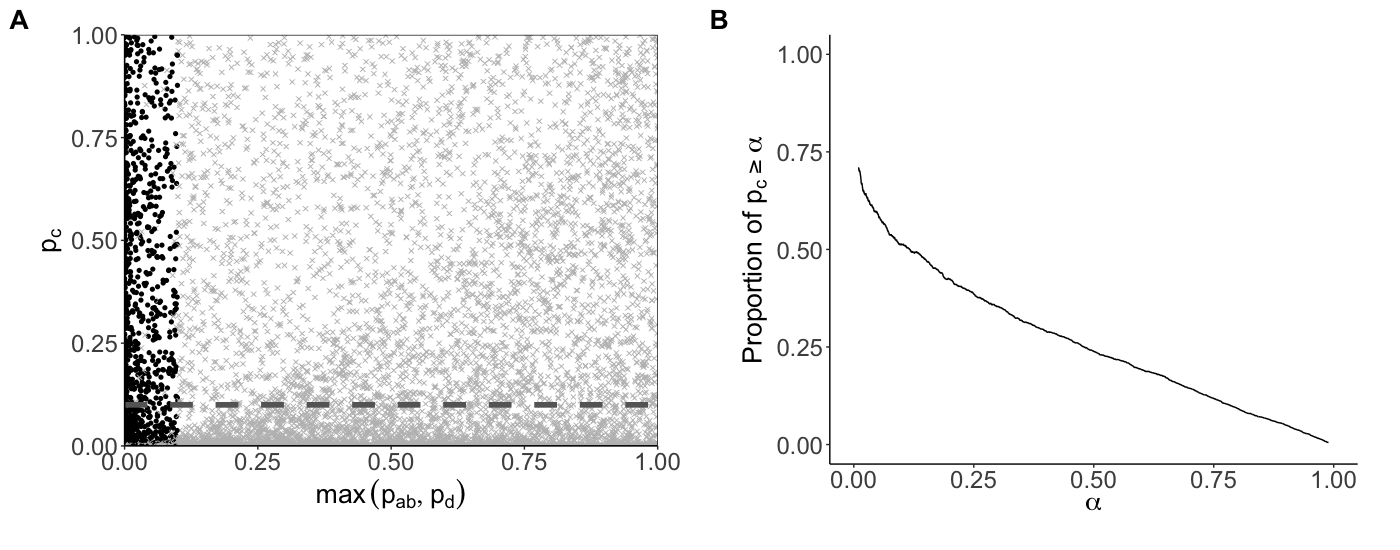}
    \caption{(A) Scatter plot of $p$-values with $\alpha = 0.1$ under LSE-Sobel framework: black solid circles represent datasets with $\max(p_{ab},p_d) < \alpha$ and $\hat{a}\hat{b}\hat{d} < 0$, grey crossings represent datasets with $\max(p_{ab},p_d) \geq \alpha$ or $\hat{a}\hat{b}\hat{d} \geq 0$,
    and the dark gray dashed line represents $p_c = \alpha$.
    (B) Proportion of datasets satisfying $p_c \geq \alpha$ for different $\alpha$ when $\max(p_{ab},p_d) < \alpha$ and $\hat{a}\hat{b}\hat{d} < 0$ under LSE-Sobel framework.}
    \label{fig: competitive_LSESobel}
\end{figure}

\begin{figure}[t!]
    \centering
    \includegraphics[scale=0.27]{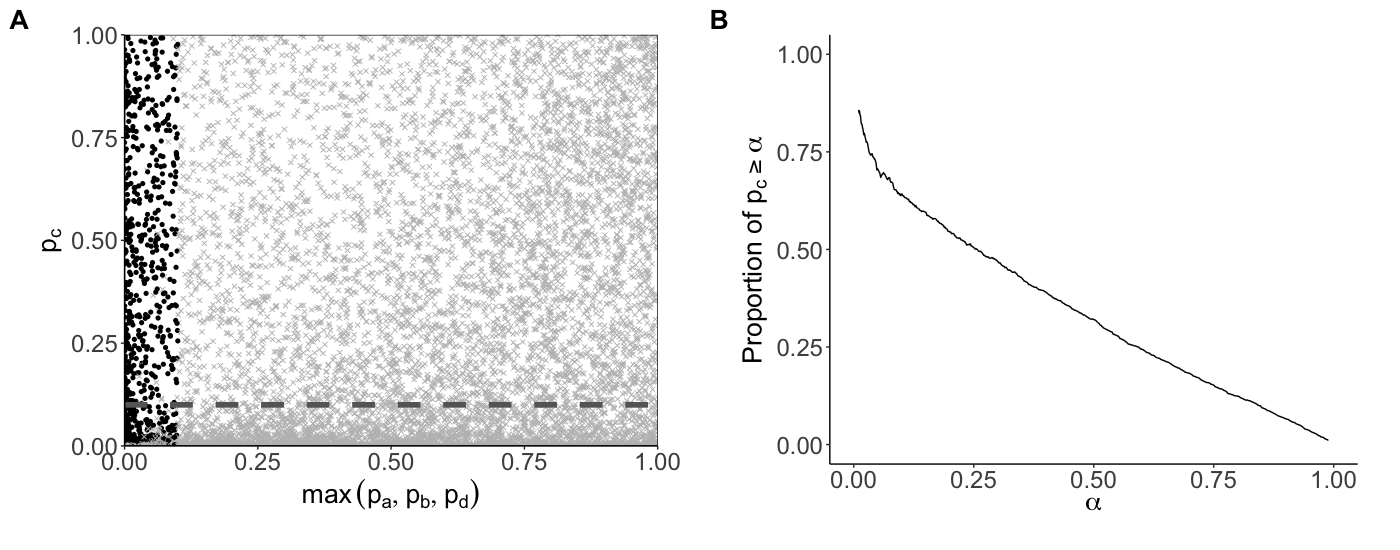}
    \caption{(A) Scatter plot of $p$-values with $\alpha = 0.1$ under LAD-$Z$ framework: black solid circles represent datasets with $\max(p_{a},p_b,p_d) < \alpha$ and $\hat{a}\hat{b}\hat{d} < 0$, grey crossings represent datasets with $\max(p_{a},p_b,p_d) \geq \alpha$ or $\hat{a}\hat{b}\hat{d} \geq 0$,
    and the dark gray dashed line represents $p_c = \alpha$.
    (B) Proportion of datasets satisfying $p_c \geq \alpha$ for different $\alpha$ when $\max(p_{a},p_b,p_d) < \alpha$ and $\hat{a}\hat{b}\hat{d} < 0$ under LAD-$Z$ framework.}
    \label{fig: competitive_LADZ}
\end{figure}

For LAD-$Z$ framework, The LAD estimator $\hat{a},\hat{b},\hat{d},\hat{c}$ as well as their corresponding $p$-values under $Z$-test are calculated. Similarly, If $\{p_c \geq \alpha\}\cap \{\hat{a}\hat{b}\hat{d}<0\} \not= \emptyset$ for any fixed $\alpha \in (0,1)$ when $\max(p_a,p_b,p_d) < \alpha$, the same conclusion can be reached under LAD-$Z$ framework.
Results are shown in Figure \ref{fig: competitive_LADZ}.

% use bibfile 
\bibliographystyle{chicago}      % Chicago style, author-year citations
\bibliography{bib}   % name your BibTeX data base

\end{document}